\documentclass[11pt,a4paper]{article}

\usepackage{./MOXReportSetting}
\pdfoutput=1
\usepackage{amsmath,amssymb,times,bm}
\usepackage{stmaryrd}
\usepackage{enumerate}
\usepackage{relsize}
\usepackage{graphicx}
\usepackage{subfigure}
\usepackage{mathrsfs}
\usepackage{latexsym}
\usepackage{amsfonts}
\usepackage{pifont}
\usepackage{pb-diagram}
\usepackage{setspace}
\usepackage{algorithm}
\usepackage{algorithmic}
\usepackage{color}
\usepackage{multirow}
\usepackage{mathtools}
\usepackage{appendix}

\DeclareMathOperator{\EX}{\mathbb{E}}
\DeclareMathOperator{\Sx}{\mathbb{S}}

\begin{document}

\newtheorem{The}{Theorem}[section]
\newtheorem{remark}{Remark}
\newtheorem{proposition}{Proposition}


\makeatletter
\newcommand{\superimpose}[2]{%
  {\ooalign{$#1\@firstoftwo#2$\cr\hfil$#1\@secondoftwo#2$\hfil\cr}}}
\makeatother

\newcommand{\abs}[1]{\left| {#1} \right|}
\newcommand{\mean}[2][]{\left\{\!\!\left\{ {#2} \right\}\!\!\right\}_{#1}}
\newcommand{\jump}[2][]{\left\llbracket {#2} \right\rrbracket_{#1}}
\newcommand{\sdot}{\mathpalette\superimpose{{\square}{\cdot}}}
\hyphenation{up-sca-ling-down-sca-ling}


\title
{
A novel dowscaling procedure for compositional data in the Aitchison geometry with application to soil texture data
}

\author
{
Federico Gatti$^{(1)}$  Alessandra Menafoglio $^{(1)}$ Niccolò Togni$^{(1)}$  Luca Bonaventura$^{(1)}$ \\
Davide Brambilla$^{(2)}$ Monica Papini$^{(2)}$ Laura Longoni$^{(2)}$
}


\maketitle
\begin{center}
{
    \small $^{(1)}$ MOX -- Modelling and Scientific Computing\\
    Dipartimento di Matematica, Politecnico di Milano \\
    Piazza Leonardo da Vinci, 20133 Milano, Italy\\
    {
        \tt
        federico.gatti@polimi.it\\
        alessandra.menafoglio@polimi.it\\
        niccolo.togni@mail.polimi.it\\
        luca.bonaventura@polimi.it\\ 
    }
    \vspace*{0.5cm}
     \small
 $^{(2)}$ Dipartimento di Ingegneria Civile e Ambientale\\
 Politecnico di Milano - Polo Territoriale di Lecco\\
 Via G. Previati 1/c, 23900 Lecco, Italy\\
 {\tt  davide.brambilla@polimi.it\\
       monica.papini@polimi.it\\ 
       laura.longoni@polimi.it\\ }
 }

\end{center}

\date{}

\vspace*{0.5cm}

\vspace*{0.5cm}

\begin{abstract}
In this work, we present a novel downscaling procedure for compositional quantities based on the Aitchison geometry. The method is able to naturally consider compositional constraints, i.e. unit-sum and positivity. We show that the method can be used in a block sequential Gaussian simulation framework in order to assess the variability of downscaled quantities. Finally, to validate the method, we test it first in an idealized scenario and then apply it for the downscaling of digital soil maps on a more realistic case study. The digital soil maps for the realistic case study are obtained from SoilGrids, a system for automated soil mapping based on state-of-the-art spatial predictions methods. 
\end{abstract}

\section{Introduction}\label{sec:intro} 

Uncertainty Quanfitifcation (UQ) is a crucial aspect for numerical tools intended to simulate physical processes, since it is important to provide an extensive analysis of the uncertainty of the outputs related to the variability of the inputs. Classical methods to perform this task are based on Monte Carlo (MC) simulations~\cite{kalos2009monte}. Here, an ensemble of realizations of the input parameters is used to feed a mathematical/numerical model, aiming to assess the distribution of the response in the face of uncertain inputs. In this broad framework, whenever parameters are characterized by a spatial distribution, geostatistical stochastic simulation can be employed to generate input scenarios for the model~\cite{brown2002stochastic}. The geostatistical approach allows one to account for the spatial dependence characterizing the input parameters and to model the spatial structure expected for the realizations (range of variability, degree of smoothness) through a spatial covariance function. Nonetheless, sound geostatistical simulation needs to take into account the possible constraints of the data, particularly when these represent compositional information. For instance, soil moisture retention plays an important role in models that simulate
hydrogeological processes and depends on a number of terrain properties, such as the \emph{soil texture}. The latter in turn is determined by particle-size fractions (psfs), i.e. the relative percentages, in terms of soil composition, of clay, silt and sand, the three categories in which grains of fine earth are divided depending on their size, see e.g.~\cite{martin:2017}. When some sparse samples are available, geostatistical techniques such as Kriging and conditional Gaussian (co)simulation can be used to interpolate the available observations and assess the associated uncertainty. However, neglecting the inherent characteristics of these data may result in inappropriate results, such as prediction of negative components or modeling spurious correlations~\cite{kim:1999}. These serious limitations hinder the use of classical geostatistical methods based on the Euclidean geometry in the presence of compositional data (see, e.g.~\cite{aitchison1982statistical, buccianti2014compositional}).

In the last decades, an increasing attention has been devoted to developing analytics tools able to account for the features of compositional data, starting from the work of Aitchison~\cite{aitchison1982statistical}. Nowadays, Compositional Data Analysis (CoDa,~\cite{egozcue:2003, pawlowsky2015modeling}) is a well-established area of statistics, which studies models and methods for compositional data, grounded on the Aitchison geometry for the simplex. The Aitchison geometry is based on the foundational idea that, in compositional vectors, only the log-ratios among components represent a meaningful information to be accounted for in the statistical analysis. In a geostatistical setting, this foundational idea led to the development of new kriging methods for compositional data, which were successfully applied in several environmental studies~\cite{pawlowsky2004geostatistical,tolosana2019geostatistics}.

In this work, we focus on the problem of geostatistical downscaling of compositional quantities. This is relevant in applications where no (or limited) direct observation is available within the study area -- because of cost or environmental constraints -- but low-resolution information is available  across the region. This is the case of our motivating study, which focuses on the stochastic characterization of soil texture within a mountain river catchment, aiming to model the hydrogeological instability -- and consequent natural hazard -- of the region. In this case, no direct observation of particle-size fractions is available, but low-resolution data are reported in public databases, such as SoilGrids~\cite{hengl:2014, hengl:2017}. In this case, characterizing the spatial distribution of the soil texture requires to operate a change of support of the available (compositional) information and to assess the corresponding uncertainty. 

To the authors' knowledge, none of the available methods for (geostatistical) downscaling allows to account for compositional constraints. For instance, methods of area-to-point kriging and stochastic simulation available in the literature~\cite{kyriakidis:2004} inevitably suffer from the limitations of the Euclidean methods. We here propose an extention of Area-To-Point Regression CoKriging (ATPRCoK) -- and associated stochastic simulation -- to compositional vectors that, based on the Aitchison geometry, allows to overcome such issues and provide stochastic scenarios for the target compositional parameters.

The remaining part of this work is organized as follows.
In Section~\ref{sec:ATPRK} we recall the area-to-point regression (co)kriging method; in Section~\ref{sec:CompositionalATPRK}, we present the downscaling prediction framework for compositional data; in Section~\ref{sec:SoilGrids} we recall the definition of psfs, which are used in Section~\ref{sec:examples} to exemplify and test the features of the method in a first to synthetic case and then in a real scenario.
Finally, we apply the method to a case study within the Caldone catchment in the Northern Italy city of Lecco, where we show how the method is able to provide psfs data at a length-scale most of the time very difficult or impossible to be determined.

\section{Area-to-point regression kriging}\label{sec:ATPRK}
In this section, we recall the main features of Area-To-Point Regression Kriging (ATPRK) and Area-To-Point Regression Cokriging (ATPRCoK); for further details see e.g.~\cite{wang2015downscaling,xiao:2018}.
Let us consider a scalar continuous random field $\{Z(\mathbf{x}), x\in D\}$ defined over a geographical region $D\subset\mathbb{R}^d$. Let us discretize $Z(\mathbf{x})$ as
\begin{equation}
Z_j = \frac{1}{|\nu_j|} \int_{\nu_j} Z(\mathbf{x}) \,d{\nu_j},
\end{equation}
where $Z_j$ denotes the discretized element at pixel $j$, $\nu_j$ defines the geographical support of the $j$-th pixel having center $\mathbf{x}_j \in D$, and $|\nu_j|$ denotes the measure of the support $\nu_j$. We assume the measure of the pixel support to be equal for all the pixels covering the region $D$ and consider two levels of spatial resolution, one coming from a coarse discretization, denoted by the index $K = 1,\dots,M$, and another coming from a fine discretization, denoted by the index $k = 1,\dots,N$. The fine and coarse supports are such that we can define an integer number $P = \frac{|\nu_K|}{|\nu_k|}$. Moreover, when using a Euclidean geometry for the data -- which is the standard setting for which ATRPK is developed -- the low-resolution random field is assumed to be obtained as an arithmetic mean of the high-resolution one, i.e., for $K=1,\dots,M,$
\begin{equation}\label{filter_eq}
Z_K = \frac{1}{P} \sum_{k: \mathbf{x}_k \in \nu_K} Z_k.
\end{equation}
Starting from one complete realization of the low-resolution field $Z_K,$ we want to estimate the high-resolution field $Z_k$, i.e. perform downscaling.
ATPRK allows one to compute an estimate of the field $Z_k$ as a (linear) combination of two parts: regression and Area-To-Point Kriging (ATPK), see e.g.~\cite{ATKINSON2013106, goovaerts:2008, kyriakidis:2004, kyriakidis2005geostatistical}. 
It uses a linear regression model on a set of covariates for the mean term of $Z_k$, and kriging to interpolate the residuals from the regression model. 
The ATPRK predictor $\widehat{Z}_k$ of the field $Z_k$ at a given fine scale pixel $\nu_k$ is defined as
\begin{equation}\label{atp_regression_kriging}
    \widehat{Z}_k =  \sum_{l} \beta^l \, u^l_{k} + \sum_{\overline{K}} \lambda^{\overline{K}} \, e_{\overline{K}},
\end{equation}
with $\beta^l, l=1,\dots,L$ and $\lambda^{\overline{K}}, \overline{K}=1,\dots, M$ unknown real quantities. The first sum in~\eqref{atp_regression_kriging} is a classical linear regression term, describing the mean of $Z_k$
\begin{equation}
    \mathbb{E}[Z_k] = \sum_{l} \beta^l \, u^l_k,
    \label{regression_trend}
\end{equation}
where $u^l_k$ are a set of known fine-resolution regressors. Given a realization of the field $Z_K$, one may linearly upscale Equation~\eqref{regression_trend} to obtain  a linear regression model for $Z_K$, i.e.,
\begin{equation}\label{linearrrrrrrrrr}
    \mathbb{E}[Z_K] = \sum_{l} \beta^l \, u^l_K,
\end{equation}
where $u_K^l$ are the upscaled regressors, i.e.
\begin{equation}\label{filter_eq_uu}
u_K^l = \frac{1}{P} \sum_{k: \mathbf{x}_k \in \nu_K} u_k^l.
\end{equation}
By combining equations~\eqref{filter_eq}-\eqref{linearrrrrrrrrr}-\eqref{filter_eq_uu}, the regression coefficents $\beta^l$ can be thus estimated by using a standard fitting procedure (e.g. Ordinary Least Squares (OLS) method) on the low-resolution field $Z_K$, see e.g.~\cite{hengl:2007, misnay:2007}. 

The second term in Equation~\eqref{atp_regression_kriging} is the Area-To-Point-Kriging (ATPK) term. It is the best linear unbiased predictor from the coarse residuals $e_K$, defined as $e_K = Z_K - \mathbb{E}[Z_K]$.  In ATPK, the residual at a given fine pixel $k$ is predicted as the best linear combination of the coarse residuals, subject to unbiasedness, i.e., $\widehat{e}_k = \sum_{\overline{K}} \lambda^{\overline{K}} \, e_{\overline{K}}$, where $\widehat{e}_k$ is the fine resolution predicted residual and $\lambda^{\overline{K}}$ solve
\begin{equation}\label{res:res}
    \min_{\lambda^{\overline{K}}\in\mathbb{R}} \mathbb{E}[(\widehat{e}_k-e_k)^2] \quad \mbox{s.t.} \quad \mathbb{E}[\widehat{e}_k] = \mathbb{E}[e_k]. 
\end{equation}
In practice, the ATPK predictor is often computed from a subset of $\overline{M}< M$ of residuals (typically selected in a neighborhood of the target pixel), to reduce the computational burden of the procedure. The optimal weights $\boldsymbol{\lambda} = (\lambda_1,\dots,\lambda_{\overline{M}})'$ are computed by minimizing the prediction error variance, which yields the following kriging linear system
\begin{equation}
    \begin{bmatrix}
    \boldsymbol{\Sigma} & \boldsymbol{1} \\
    \boldsymbol{1}^T & 0
    \end{bmatrix}
    \begin{bmatrix}
    \boldsymbol{\lambda} \\
    \mu 
    \end{bmatrix}
    = 
    \begin{bmatrix}
    \boldsymbol{\sigma} \\
    1 
    \end{bmatrix}.
    \label{kriging_system}
\end{equation}
Here, the element in position $(\overline{K}_1,\overline{K}_2)$ of matrix $\boldsymbol{\Sigma}$ is the block-block covariance between the residuals at the coarse pixels centered at $\mathbf{x}_{\overline{K}_1}$ and $\mathbf{x}_{\overline{K}_2}$; the $\overline{K}$-th element of $\boldsymbol{\sigma}$ is the point-block covariance of the residuals between fine and coarse pixels respectively centered at $\mathbf{x}_k$ and $\mathbf{x}_{\overline{K}}$, and $\mu$ is a Lagrange multiplier.

Note that, in practice, neither the residuals nor their covariance are observed, but need to be estimated from the data. Residuals are typically estimated by difference from the estimated regression term. Estimating the covariance structure is more critical.
Under the assumption that the residual field $e_k$ is stationary and isotropic and denoting with $C_{k_1,k_2}$ the covariance between pixels $k_1$ and $k_2$ of the residual at fine scale, we can compute the block-block covariance as
\begin{align}
    \label{areal_covariance}
    \begin{split}
        \Sigma_{\overline{K}_1,\overline{K}_2} = \frac{1}{P^2} \sum_{i=1}^{P} \sum_{j=1}^{P} C_{i,j}, \quad \mathbf{x}_i\in \nu_{\overline{K}_1}, \mathbf{x}_j\in \nu_{\overline{K}_2}.
    \end{split}
\end{align}
The point-block covariance is then given by
\begin{align}
    \label{area_point_covariance}
    \begin{split}
        \sigma^{\overline{K}} = \frac{1}{P} \sum_{i=1}^{P} C_{i,k}, \quad \mathbf{x}_i\in \nu_{\overline{K}}.
    \end{split}
\end{align}

The critical point of the ATPK method is thus the determination of the covariance structure at the fine scale, which cannot be directly estimated, as the data are given at the coarse scale only. The problem of estimating the fine-scale semi-variogram $\gamma_{k_1,k_2}$
\begin{equation}
    \gamma_{k_1,k_2} = C_{k_1,k_1}-C_{k_1,k_2} = \frac{1}{2} \, \mathbb{E}[(e_{k_2} - e_{k_1})^2],
\end{equation}
from coarse-scale data is known as a \emph{deconvolution problem}. We shall not focus on this problem in this work, as it is completely analogous to that encountered in the Euclidean setting -- for which a number of methods are available. We refer to~\cite{goovaerts:2008} for more details on the deconvolution method used in this work. 

\smallskip
In case of multi-dimensional random fields, the ATPRK framework changes slightly in order to take into account possible cross-correlations among field components. Generalization of ATPRK to the multivariate case is analogous to cokriging, and yields Area-To-Point-Regression-CoKriging (ATPRCoK), see e.g.~\cite{xiao:2018}. In ATPRCoK the coarse residuals appearing in \eqref{atp_regression_kriging} are replaced by the residuals of all the components of the multi-dimensional random field, in order to consider possible cross-correlations among their components. If we consider a $p$-dimensional random field $\{\mathbf{Z}(\mathbf{x}),$ $\mathbf{x} \in D\},$ its ATPRCoK discrete prediction is,
\begin{equation}
    \widehat{\mathbf{Z}}_k = \sum_l u_k^l \boldsymbol{\beta}^l + \sum_{\overline{K}} \Lambda_{\overline{K}} \mathbf{e}_{\overline{K}},
\end{equation}
where $\boldsymbol{\beta}^l \in \mathbb{R}^{p}$ are the vectors of the unknown regression coefficients. The matrix $\Lambda_{\overline{K}} \in \mathbb{R}^{p\times p}$ contains the unknown cokriging coefficients and $\mathbf{e}_{\overline{K}} \in \mathbb{R}^p$ is the vector of the coarse scale residuals. The optimal weights $\Lambda_{\overline{K}}$ are found by solving a system analogous to \eqref{kriging_system}, but considering covariances and cross-covariances within/among fields components, as in a standard cokriging setting.

\section{Compositional ATPRCoK}\label{sec:CompositionalATPRK}
In this section, we consider the problem of downscaling compositional data and we propose a method which extends ATPRCoK to the Aitchison geometry, and naturally takes into account the compositional nature of the data.

\subsection{Compositional data in the Aitchison simplex}
A compositional data point $\mathbf{Z} = (Z_1,...,Z_p)$ is typically represented as a vector whose elements are proportions (or percentages) of a whole, named \emph{total}. In this case, compositional vectors are characterized by the unit-sum constraint $\sum_{i} Z_i = 1,$ where we denote with $Z_i \ge 0$ the $i$-th component of compositional data point. More generally, compositional vectors are data which convey only relative information, being subject to a constant-sum constraint. Here the total is typically of no interest for the analysis, in the sense that expressing the data w.r.t. a different total (i.e., in proportion, percentages or ppm) should not change the results of the analysis (i.e., \emph{scale invariance}).
Because of the range limitation and the possible spurious correlation of compositional vectors~\cite{kim:1999, pawlowsky1984spurious}, the Euclidean-based statistical framework was proved to be ineffective for the spatial prediction of this type of data, although a number of authors have ignored this aspect, see e.g.~\cite{delbari:2011}. Other works (e.g.~\cite{walvoort:2001}) tried to account for the particular nature of regionalised variables expressing relative fractions by proposing an extension of kriging called Compositional Kriging (CK). CK predictions respect the constraints of positivity and constant sum value. However, the CK algorithm is based on empirical considerations rather than a coherent probabilistic model, and is therefore not suited for stochastic simulation. Our developments follow the direction of research on compositional kriging explored by~\cite{pawlowsky1989cokriging, tolosana2011geostatistics}, who formulated geostatistical models and methods based on the Aitchison geometry for the simplex (see~\cite{pawlowsky2016spatial, tolosana2019geostatistics} for recent reviews).

Presently, the standard approach to the statistical analysis of compositional data is the one pioneered by Aitchison, see~\cite{aitchison1982statistical}, which is based on the particular geometry of the simplex~\cite{aitchison:1986, pawlowsky2015modeling, pawlowsky2001geometric}. A $p$-dimensional compositional vector $\mathbf{Z} = (Z_1,\dots,Z_p)$ is an element of the $p$-dimensional standard simplex, $\mathbb{S}^p$, which is defined as
\begin{equation}\label{nsimplex}
    \mathbb{S}^p = \{(Z_1,\dots,Z_p: Z_i \ge 0, \: \sum_{i = 1}^p Z_i = 1\}.
\end{equation}

In~\cite{aitchison:1986, pawlowsky2001geometric} group operations are defined to give the simplex a structure of a real vector space. These are the perturbation $\oplus$ (sum) and powering $\odot$ (product by a constant) operations, defined, for $\mathbf{x},\mathbf{y} \in \mathbb{S}^p$, and $\alpha\in\mathbb{R}$, respectively as
\begin{align}
    \begin{split}
        \mathbf{x} \oplus \mathbf{y} = \mathcal{C}\left(x_1 y_1,\dots, x_{p} y_{p} \right),
         \\
        \alpha \odot \mathbf{x} = \mathcal{C}\left(x_1^\alpha,\dots, x_{p}^\alpha\right).
    \end{split}
\end{align}
Here, $\mathcal{C}(\cdot)$ denotes the closure operation
\begin{equation}
\mathcal{C}(\mathbf{x}) = \left(\frac{x_1}{\sum_{i=1}^p x_i},\dots,\frac{x_p}{\sum_{i=1}^p x_i}\right)\ \ \mathbf{x} \in \mathbb{R}_+^p.
\end{equation}

The space $\mathbb{S}^p$ can be equipped with a (finite-dimensional) Hilbert space structure when considering the Aitchison inner product, defined, for $\mathbf{x},\mathbf{y} \in \mathbb{S}^p$ as
\begin{equation}
\langle \mathbf{x}, \mathbf{y} \rangle_a = \frac{1}{2p} \sum_{i=1}^{p} \sum_{j=1}^{p} \ln \frac{x_i}{x_j} \cdot \ln \frac{y_i}{y_j}\ \ \mathbf{x}, \mathbf{y} \in \mathbb{S}^p.
\end{equation}

\noindent The defined inner product induces a norm $\| \cdot \|_{a} := \sqrt{\langle \cdot,\cdot \rangle_a},$ which in turn induces a distance $d_{a}(\mathbf{x}, \mathbf{y}) = \|\mathbf{x} \ominus \mathbf{y}\|_{a}, \ \mathbf{x}, \mathbf{y} \in \mathbb{S}^p,$ where $\mathbf{x} \ominus \mathbf{y}$ denotes the perturbation of $\mathbf{x}$ with the reciprocal of $\mathbf{y},$ i.e., $\mathbf{x} \ominus \mathbf{y} = \mathbf{x} \oplus ( (-1) \odot \mathbf{y} ).$ 
The Hilbert space structure identified by these operations is called Aitchison geometry, or Aitchison simplex~\cite{pawlowsky2001geometric}.

\subsection{ATPRCoK in the Aitchison geometry}

The statistical approach proposed by Aitchison~\cite{aitchison1982statistical} and following authors~\cite{pawlowsky2015modeling,pawlowskiglahn:2001} consists in analyzing compositional data in the context of the Aitchison geometry. Here, a large number of multivariate statistical methods grounded on a geometric perspective (e.g., principal component analysis. regression) can be properly reformulated to account for the inherent properties of compositional data. From an operation viewpoint, the standard procedure of analysis consists in transforming the original data by applying an isomorphism from the $p$-dimensional Aitchison simplex to the classical Euclidean space $\mathbb{R}^{p-1}$ or $\mathbb{R}^{p}$, perform the statistical analysis on the transformed data and finally back-transform the results in the original space. This strategy was proved to be fully equivalent to working directly in the Aitchison simplex for a number of statistical methods (see, e.g.~\cite{egozcue:2003}). 
In this section, we shall formulate ATPRCoK in the Aitchison simplex, and then prove that one may equivalently perform the analysis by relying on the so-called Isometric Log-Ratio (ILR) transformation, which is an isometry that maps the simplex in $\mathbb{R}^{p-1}.$ The latter associates to a compositional vector $\mathbf{z} \in S^p$ the coordinates of this vector with respect to an orthonormal basis of the simplex $(\boldsymbol{\psi}_1,\dots,\boldsymbol{\psi}_{p-1})$, i.e., 
    \begin{equation}
        \text{ILR}(\mathbf{z}) = \left( \left< \mathbf{z}, \boldsymbol{\psi}_1 \right>_a,\dots,\left< \mathbf{z}, \boldsymbol{\psi}_{p-1} \right>_a \right)^{'}.
        \label{ILR} 
    \end{equation}
Note that the ILR is a linear transformation, see e.g.~\cite{egozcue:2003}. For several compositional methods (e.g., principal component analysis, regression, see, e.g.,~\cite{pawlowsky2015modeling}), it was shown that the choice of the basis does not influence the results of the analysis. However, specific choices for the basis can lead to practical advantages. For instance, the basis could be chosen in such a way as to grant uncorrelation of the resulting transformed data, or to ease the interpretation of the results (see, e.g.~\cite{fivserova2011interpretation}). 

In~\cite{dobarco:2016} the authors used ATPCoK to downscale psfs data transformed with Additive-Log Ratio (ALR), using the silt fraction as a reference for the ratios. ALR was there used as a practical solution to account for the compositional nature of the data, but the modelling assumptions were not explicitly stated, and the results were interpreted only in terms of prediction accuracy with respect to a given test set. 
Recent works highlighted some limitation of the ALR transformation, as this is not isometric, thus does not preserve the Aitchison geometry (see, e.g., \cite{pawlowsky2015modeling}). In this section, we propose a general method for the statistical downscaling and simulation of compositional data which extends the ATPRCoK to the context of the Aitchison simplex. 
We call the method ILR-ATPRCoK to recall the computational strategy we propose to perform ATPRCoK in the Aitchison simplex, which is based on the ILR transformation.  Here, we shall also prove that the strategy based on ILR is fully equivalent to working directly in the simplex itself.  

In the following, we denote by $\{\mathbf{Z}(\mathbf{x}), \mathbf{x} \in D\}$ a random field valued in $\mathbb{S}^p$, defined over a Euclidean region $D\subset\mathbb{R}^d$. To indicate the Aitchison center of the field (i.e. the mean value in Aitchison geometry), the Aitchison covariance and the integral operator of $\mathbf{Z}(\mathbf{x})$ over a region $\tau \subset \mathbb{R}^d$, we use respectively the same notation used in~\cite{menafoglio:2014,pawlowsky2011compositional}, that is
\begin{itemize}
  \item Aitchison center,
  \begin{equation}
    \boldsymbol{\mu}(\mathbf{x}) = Cen(\mathbf{Z}(\mathbf{x})) = \text{arg} \text{min}_{\mathbf{z} \in \Sx^p} \EX[d^2_a(\mathbf{Z}(\mathbf{x}), \mathbf{z})];
  \end{equation}
   \item Aitchison covariance operator, acting on a (non-random) element $\mathbf{z}\in\mathbb{S}^p$, for $\mathbf{x}_1, \mathbf{x}_2 \in D$, as
   \begin{equation}
     C_{a}(\mathbf{x}_1,\mathbf{x}_2)\mathbf{z} = Cen[\langle \mathbf{Z}(\mathbf{x}_1) \ominus \boldsymbol{\mu}(\mathbf{x}_1), \mathbf{z} \rangle_{a} \odot (\mathbf{Z}(\mathbf{x}_2) \ominus \boldsymbol{\mu}(\mathbf{x}_2))];
   \end{equation}
  \item Aitchison integral
  \begin{equation}
  \int_{\tau}^{\oplus}\mathbf{Z}(\mathbf{x})d\tau = \mathcal{C}\left( \  e^{\int_{\tau}\ln(Z_{1}(\mathbf{x}))d\tau},..., e^{\int_{\tau}\ln(Z_{p}(\mathbf{x}))d\tau} \ \right).
  \label{integral:aitchison}
  \end{equation}
\end{itemize}
We refer to~\cite{aitchison:1986, egozcue:2003} for an insight of the geometry of the random compositions in the Aitchison simplex, and to \cite{bosq2000} for a recall on covariance operators in Hilbert spaces.

For the element $\mathbf{Z}(\mathbf{x})$ of the compositional field at $\mathbf{x}\in D$, we assume the following model
\begin{equation}
    \mathbf{Z}(\mathbf{x}) = \boldsymbol{\mu}(\mathbf{x})  \oplus \mathbf{e}(\mathbf{x}),
    \label{mod_Z_compositional}
\end{equation}
with $\mathbf{e}(\mathbf{x})$ the residual term. We model the center of the field through a linear model in $\mathbb{S}^p$
\begin{equation}
    \boldsymbol{\mu}(\mathbf{x}) = \bigoplus_l u^l(\mathbf{x}) \odot \boldsymbol{\beta}^l,
    \label{mod_drift}
\end{equation}
where $\boldsymbol{\beta}^l \in \mathbb{S}^{p}, l=1,\dots,L$ are the vectors of unknown regression coefficients and $u^l(\mathbf{x}) \in \mathbb{R}, \mathbf{x} \in D,$ are the covariates. Furthermore, we assume that the residual is second-order stationary, see~\cite{tolosana2019geostatistics}, i.e., that the covariance structure (in Aitchison geometry) for a random composition $\mathbf{Z}(\mathbf{x}), \mathbf{x} \in D$, only depends on the increment among locations
\begin{equation}
    C_{a}(\mathbf{x}_1,\mathbf{x}_2) = \widetilde{C}_{a}(\mathbf{x}_1 - \mathbf{x}_2), \ \ \mathbf{x}_1, \mathbf{x}_2 \in D.
\end{equation}
To simplify the notation, we shall indicate the stationary covariance function $\widetilde{C}_{a}$ simply by $C_{a}$. 

With a notation analogue to that used in Section \ref{sec:ATPRK} we consider the discretized versions of the field $\{\mathbf{Z}(\mathbf{x}), \mathbf{x}\in D\}$, denoted by $\mathbf{Z}_k$ (resp. $\mathbf{Z}_K$) and obtained at a fine (resp. coarse) discretization scale, namely
\begin{equation}
\mathbf{Z}_k = \frac{1}{|\nu_k|} \odot \int_{\nu_k}^\oplus \mathbf{Z}(\mathbf{x})d\nu_k, \quad 
\mathbf{Z}_K = \frac{1}{|\nu_K|} \odot \int_{\nu_K}^\oplus \mathbf{Z}(\mathbf{x})d\nu_K.
\end{equation}
Here, the powering by $\frac{1}{|\nu_k|}$ and $\frac{1}{|\nu_K|}$ is intended as acting element-wise. 

Given the realization of the coarse-scale field $\mathbf{Z}_K$, and by analogy with \eqref{atp_regression_kriging}, we define the ILR-ATPRCoK predictor of the fine-scale field $\mathbf{Z}_k \in \mathbb{S}^p$ as
\begin{equation}
    \widehat{\mathbf{Z}}_k = \bigoplus_l u^l_k \odot \boldsymbol{\beta}^l  \oplus \bigoplus_{K} \Lambda_K \boxdot \mathbf{e}_K,
    \label{area_to_point_kriging_estimate_aitchison}
\end{equation}
where $\Lambda_K \in \mathbb{R}^{p \times p}$ is a matrix of ATPCoK unknown weights to be optimized, $\boxdot$ is the matrix-by-composition multiplication -- consistent with perturbation and powering, as defined in \cite{pawlowsky2015modeling} (p. 55), i.e., denoting by $e_{K,i}, i=1,\dots,p$ the elements of $\mathbf{e}_K\in \mathbb{S}^p$ and by $\lambda_{K,i,j}, i,j=1,\dots,p$ the elements of $\Lambda_K$
\begin{align*}
    \Lambda_K \boxdot \mathbf{e}_K = \mathcal{C}\left[\prod_{j=1}^p e_{K,j}^{\lambda_{K,1,j}},\dots,\prod_{j=1}^p e_{K,j}^{\lambda_{K,p,j}}\right].
\end{align*}

The residuals $\mathbf{e}_K \in \mathbb{S}^p$ represent the upscaled residuals of \eqref{mod_Z_compositional}, defined as
\begin{equation}\label{resid_aitch}
\mathbf{e}_K = \frac{1}{|\nu_K|} \odot \int_{\nu_K}^\oplus (\mathbf{Z}(\mathbf{x}) \ominus \boldsymbol\mu(\mathbf{x}))d\nu_K.
\end{equation}

Under the assumption that the regression coefficients $\boldsymbol{\beta}^l$ and the covariance function $C_a$ are known, the optimal weights $\Lambda_k$ in~\eqref{area_to_point_kriging_estimate_aitchison} are found as to guarantee that the ILR-ATPRCoK is the Best Linear Unbiased Predictor (BLUP) in $\mathbb{S}^p$, i.e., by solving the constrained minimization problem 
\begin{equation}
\arg\min_{\Lambda_K \in \mathbb{R}^{p\times p}}\:\mathbb{E}\left[d_a^2\left( \bigoplus_{K} \Lambda_K \boxdot \mathbf{e}_K, \mathbf{e}_k\right)\right]\quad \text{s.t.}\quad  Cen\left(\bigoplus_{K} \Lambda_K \boxdot \mathbf{e}_K\right) = 
\overline{\boldsymbol{\mu}},
\label{ILR-ATPK-obj}
\end{equation}
where $\overline{\boldsymbol{\mu}}$ is the spatially constant residual mean. 

The following result states that, by applying the ILR transformation (i.e. mapping from the Aitchison simplex to an Euclidean space, via an isometric isomorphism), one obtains an equivalent formulation of problem \eqref{ILR-ATPK-obj} that can be solved using the standard ATPRCoK presented in Section~\ref{sec:ATPRK}. 
The proof of Proposition~\ref{PropositionNewComp} is given in Appendix~\ref{appendixProof}.

\begin{proposition}\label{PropositionNewComp}
Given a compositional random field $\mathbf{Z}(\mathbf{x})$ valued in $\mathbb{S}^p$ and a random field $\mathbf{Y}(\mathbf{x})$ valued in $\mathbb{R}^{p-1}$ defined as $\mathbf{Y}(\mathbf{x}) = \text{ILR} (\mathbf{Z}(\mathbf{x}))$ for $\mathbf{x} \in D$, the BLUP in $\mathbb{S}^p$ for $\mathbf{Z}_k$ -- found by solving \eqref{ILR-ATPK-obj} --  coincides with the ILR-back-transformed ATPRCoK predictor for $\mathbf{Y}_k$ defined in \eqref{atp_regression_kriging}, i.e.,
\begin{equation}
  \widehat{\mathbf{Z}}_k = \text{ILR}^{-1} (\widehat{\mathbf{Y}}_k ).
 \label{predictors:equivalence}
\end{equation}
\end{proposition}

Even though $\boldsymbol{\beta}^l$ is rarely \emph{a priori} known, an estimate of  $\boldsymbol{\beta}^l$ can be obtained by back-transforming the corresponding estimate of the coefficient vectors $\boldsymbol{\beta}^l_{\mathbf{Y}}$ referred to the ILR-transformed field $\mathbf{Y}(\mathbf{x})$ (see, e.g., \cite{pawlowsky2015modeling}). Similarly, an estimate of the covariance operator $C_a$ can be obtained from the estimated (Euclidean) covariance operator $C_\mathbf{Y}$ of the vector field $\mathbf{Y}(\mathbf{x})$. In this work, for $\boldsymbol{\beta}^l_{\mathbf{Y}}$ we shall consider OLS estimates, whereas for $C_\mathbf{Y}$ the estimates obtained by Goovaerts' deconvolution \cite{goovaerts:2008} of classical cross-variograms.

Note that the equivalence between the ATPRCoK in the Aitchison simplex and the Euclidean ATPRCoK on ILR-transformed data (as stated in Prop.~\ref{PropositionNewComp}), implies the possibility to analogously perform Block Sequential Gaussian Simulation (BSGS)~\cite{soares:2001}, as BSGS grounds on the same hypothesis as ATPRCoK, and it is indeed based on the latter method. In the context of Uncertainty Quantification (UQ), BSGS is key to propagate the uncertainty in numerical models that take as input downscaled compositional data, as we discuss in Section~\ref{sec:CaldoneCaseStudy}.

Finally, one should note that, since the assumptions are made with respect to the Aitchison geometry, the \textit{mass-preserving} property as stated in the Euclidean framework, i.e. (see~\cite{kyriakidis:2004}), 
\begin{equation}
     \mathbf{Z}_{K} = \frac{1}{P} \sum_{\mathbf{x}_k \in \nu_K} \widehat{\mathbf{Z}}_k,
\end{equation}
does not hold. In a discrete prediction setting, as in Section~\ref{sec:ATPRK}, the Aitchison geometry predictions respect the following \textit{centre-preserving} property
\begin{equation}
     \mathbf{Z}_{K} = \frac{1}{P} \odot \bigoplus_{\mathbf{x}_k \in \nu_K} \widehat{\mathbf{Z}}_k.
    \label{centre_preserving_prediction}
\end{equation}
Indeed, since $P = \frac{|\nu_K|}{|\nu_k|} \in \mathbb{Z}^+$, as defined in Section~\ref{sec:ATPRK}, one has
\begin{equation}
    \mathbf{Z}_{K} = \mathcal{C}\left( \left(\prod_{k=1}^{P} \widehat{Z}_{1,k} \right)^{\frac{1}{P}},...,\left(\prod_{k=1}^{P} \widehat{Z}_{n,k} \right)^{\frac{1}{P}}\right).
    \label{centre_preserving_aitchison}
    \vspace{0.5cm}
\end{equation}
This means that, in the Aitchison simplex, coarse areal data coincide with the geometric mean of the predicted fine areal values (normalized to having unit-sum).

In the following sections we exemplify the proposed methodology through its application to particle-size fractions, whose definition is recalled in the next Section \ref{sec:SoilGrids}.

\section{Particle size fractions}\label{sec:SoilGrids}
Soil texture is a classification instrument used to determine soil classes. More specifically, soil texture is quantitatively determined on the basis of the relative fractions of the fine particles of different sizes that compose the terrain. Soil particles under $2\:mm$ are divided in three groups
\begin{itemize}
    \item clay: particles with a diameter less than $2\:\mu m$;
    \item silt: particles with a diameter comprised between $2\:\mu m$ and $50\:\mu m$;
    \item sand: particles with a diameter comprised between $50\:\mu m$ and $2\:mm.$
\end{itemize}
Fractions of clay, silt and sand are usually indicated as particle-size fractions (psfs). Soil texture classes are determined by the relative percentages of psfs, according to a standard that may vary depending on the country.
\begin{figure}[htbp]
  \centering
  \includegraphics[scale = 0.3]{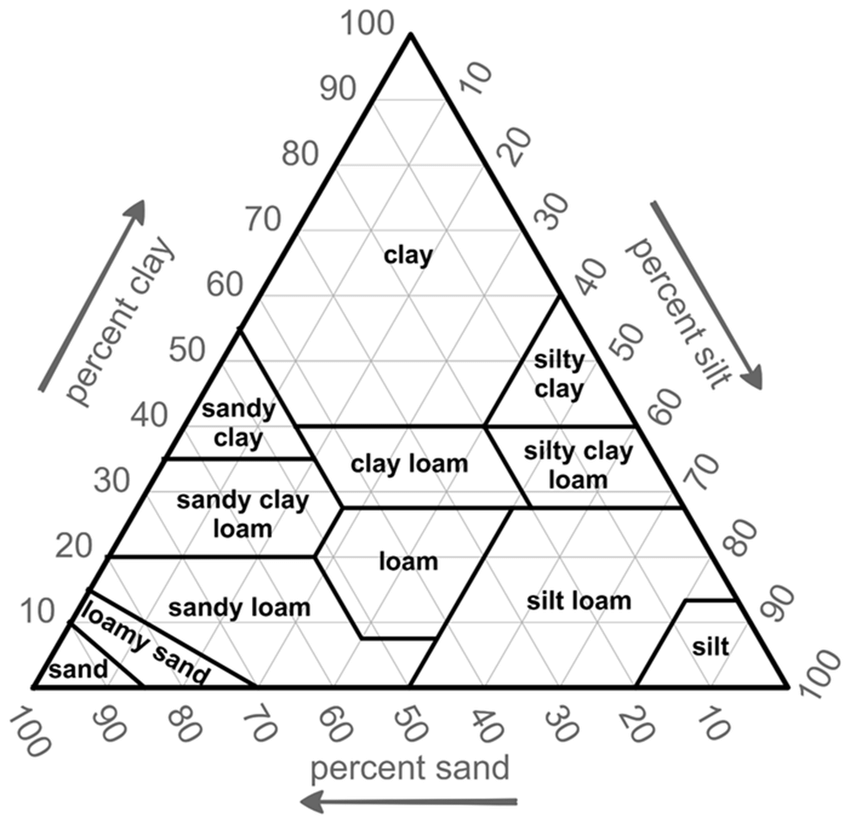}
  \caption{Soil texture triangle: Soil texture classification according to the USDA classification system, based on relative fractions of clay, silt and sand. Figure taken from \cite{USDAfigure}.}
  \label{fig:triangle}
\end{figure}
The most common classification is that used by the United States Department of Agriculture (USDA)~\cite{schaefer2007usda}, which distinguishes twelve major soil texture classes shown in Figure~\ref{fig:triangle}. The classes are typically named after the primary constituent particle-size or a combination of the most abundant particles sizes, e.g. sandy clay or silty clay. A fourth term, ``loam'', is used to describe equal proportions of sand, silt, and clay in a soil sample, and leads to the naming of even more classes, e.g. clay loam or silt loam.

\section{Validation}\label{sec:examples} 
The geostatistical method outlined in the previous sections has been implemented in R-3.6~\cite{r:2008} using the libraries \textit{gstat}~\cite{gstat:2016, gstat:2004} for ATPK and geostatistical simulation and \textit{compositions}~\cite{vanderboogaart:2008} for the analysis of compositional data. In particular, for the variogram deconvolution we use the Goovaerts' procedure~\cite{goovaerts:2008, goovaerts:2010}. We define a continuous random field $\mathbf{Z}(\mathbf{x})=(Z_1(\mathbf{x}),Z_2(\mathbf{x}),Z_3(\mathbf{x}))$ where

\begin{align*}
    \begin{split}
        Z_1(\mathbf{x}) &= \text{$\%$ of part 1 at location $\mathbf{x}$ of the domain $D;$} \\
        Z_2(\mathbf{x}) &= \text{$\%$ of part 2 at location $\mathbf{x}$ of the domain $D;$} \\
        Z_3(\mathbf{x}) &= \text{$\%$ of part 3 at location $\mathbf{x}$ of the domain $D.$}
    \end{split}
\end{align*}

In our analysis, compositional data are transformed using the function ILR of the package \textit{compositions}, see e.g.~\cite{vanderboogaart:2008}. The basis used for the transformation is the one introduced in the original article \cite{egozcue:2003}, based on the partition of the vector of compositional variables in two sub-compositions, the first consisting of $Z_{1}$ and $Z_{2}$ and the second containing only $Z_{3}$. For the sake of illustration and in view of the motivating study, we shall interpret $\mathbf{Z}(\mathbf{x})$ as the psfs at $\mathbf{x}$ (i.e., $Z_1(\mathbf{x})$, $Z_2(\mathbf{x})$, $Z_3(\mathbf{x})$ represent the composition in clay, silt and sand, respectively). Nonetheless, the validity of the simulation study here presented is clearly not limited to the specific case of psfs.

\subsection{Synthetic data}\label{SyntheticDataAnalysis}
\begin{figure}[tbp]
    \centering
    \includegraphics[scale=.25]{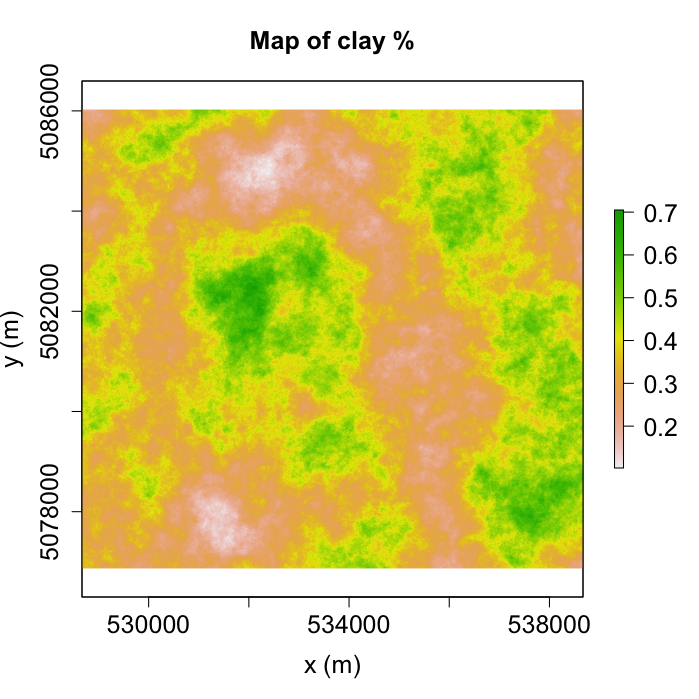}\hfill
    \includegraphics[scale=.25]{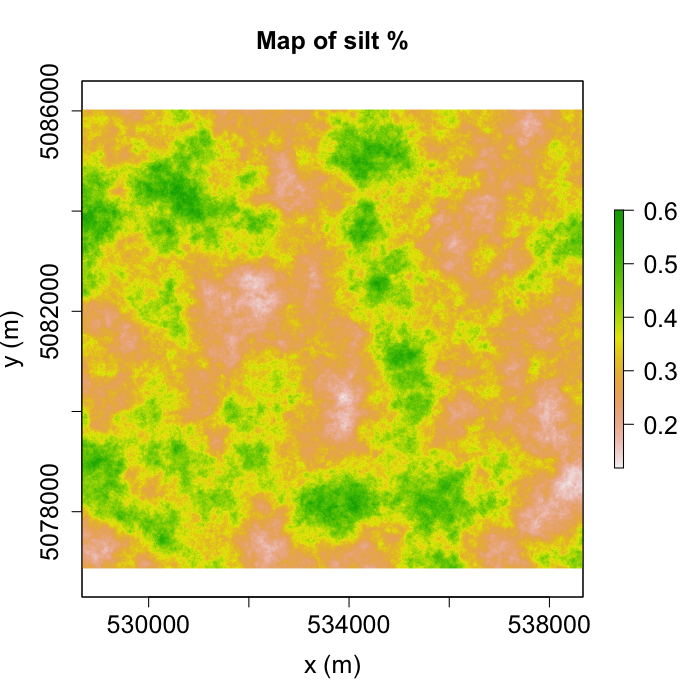}\vfill
    \includegraphics[scale=.25]{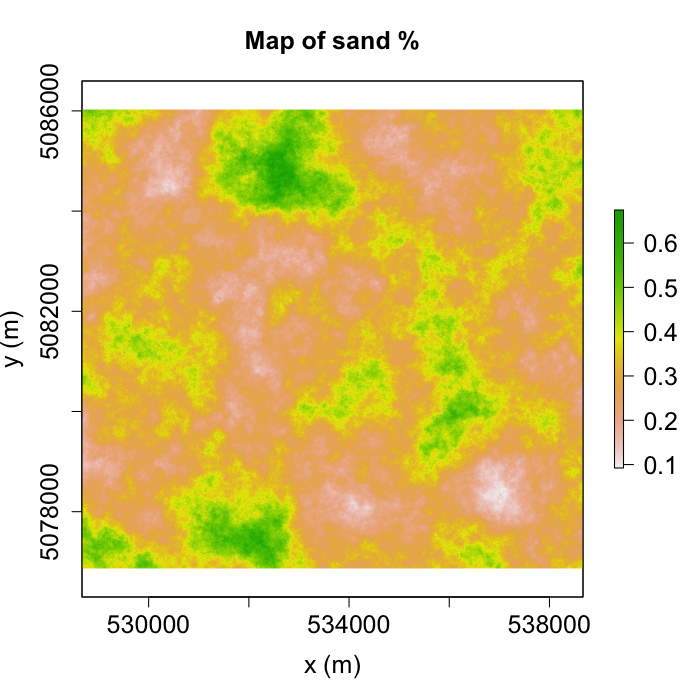}\hfill
    \includegraphics[scale=.25]{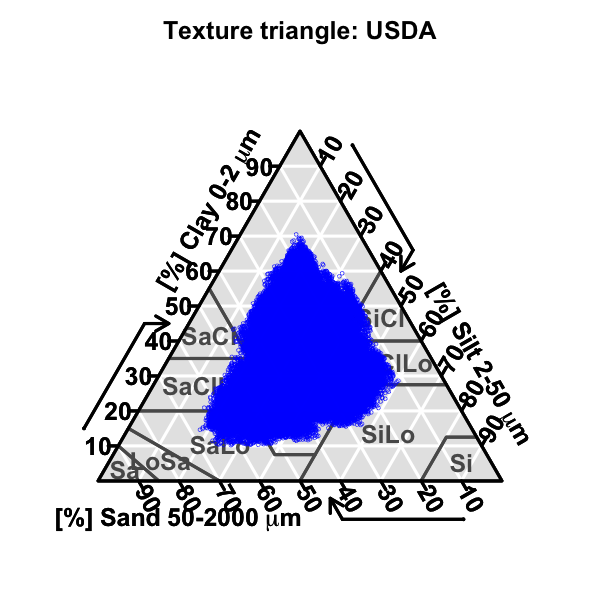}
    \caption{One realization of the initial psfs field $\mathbf{Z}(\mathbf{x})$ with $Cen(\mathbf{Z}(\mathbf{x}))=(\frac{1}{3},\frac{1}{3},\frac{1}{3})$ and $\sigma^2=0.1.$}
    \label{fig:idealizedTriangle}
\end{figure}

To assess the performance of the proposed method, we consider a simulated dataset $\mathbf{Z}(\mathbf{x}), \mathbf{x} \in D$, with support measure $|\nu_k| = 20\times20\:m^2$, on a given rectangular domain $D$ with area $|D| = 10000\times9160\:m^2$. The compositional vector $\mathbf{Z}(\mathbf{x})$ is modelled as a process with a given spatially-constant center $Cen(\mathbf{Z}(\mathbf{x})) = \boldsymbol{\mu}$ and stationary-isotropic covariance structure. From the operational viewpoint, the mean $\boldsymbol{\mu} = \mathcal{C}[({\mu}_1, {\mu}_{2}, \mu_3)']$ is set based on independent uniform distributions $\mu_i \sim U[0,1], i=1,2,3$. Compositions were simulated by back-transforming through $ILR^{-1}$ two-dimensional Gaussian random vectors $\mathbf{Y}$, with constant mean $\boldsymbol{\mu}_\mathbf{Y}=ILR(\boldsymbol{\mu})$, and stationary-isotropic marginal variograms from the spherical model without nugget ~\cite{chiles2009geostatistics, cressie1992statistics}. For the following simulations, the components of $\mathbf{Y}$ are always assumed to be uncorrelated, and the marginal ranges are both set to $2000\:m$. In each simulation, the common sill $\sigma^2$ is sampled according to a uniform distribution $U[0.025, 2.5]$. In Figure~\ref{fig:idealizedTriangle} we show an example of realization of the psfs distribution.

Starting from this set of synthetic psfs, we perform a sequence of upscaling-down\-scaling procedures, as follows. Downscaling is done using either ATPRCoK or ILR-ATPRCoK and upscaling either in Euclidean or Aitchison geometry, so that four different possibilities arise. In the following, we call AA the upscaling in the Aitchison simplex and downscaling via ILR-ATPRCoK, EE the upscaling in the Euclidean space and downscaling via ATPRCoK whereas EA, AE are the mixed methods, referring respectively to upscaling in the Euclidean space and to downscaling via ILR-ATPRCoK and upscaling in the Aitchison geometry and downscaling via ATPRCoK.

\begin{figure}[htbp]
    \centering
    \includegraphics[scale=.25]{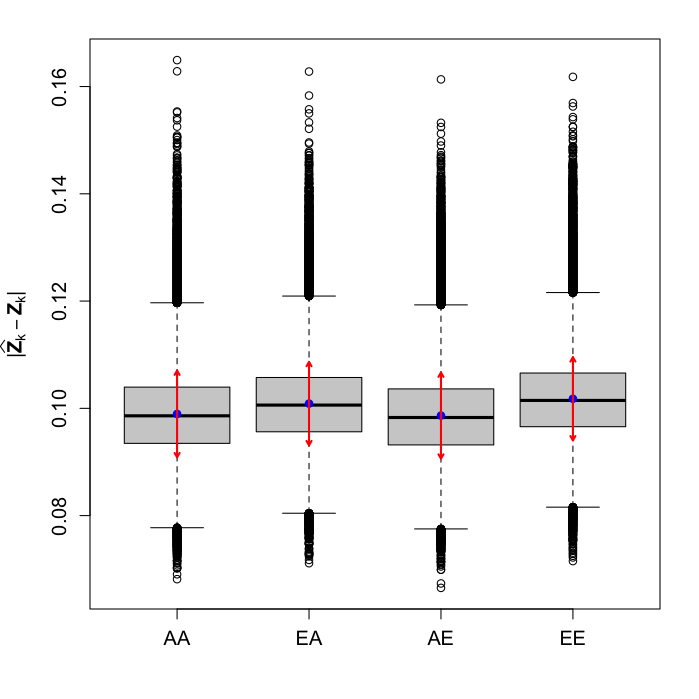}a)\\
    \includegraphics[scale=.25]{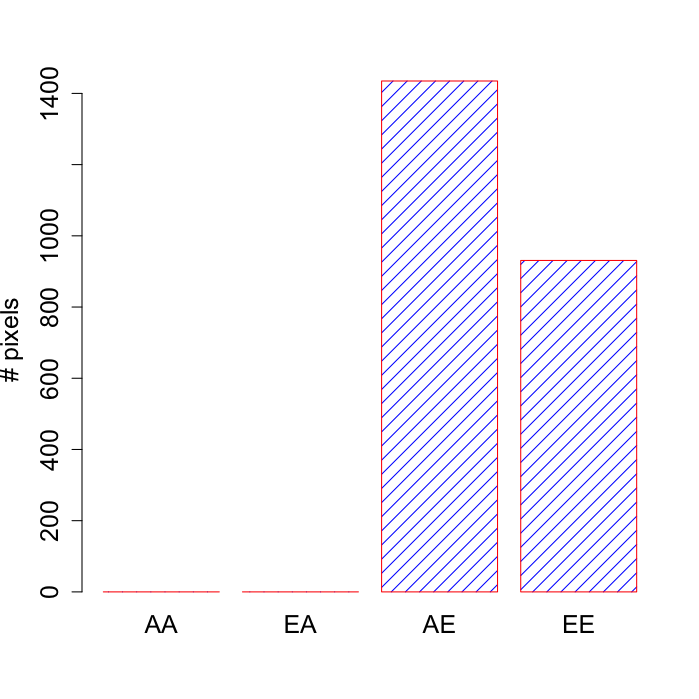}b)\\
    \includegraphics[scale=.25]{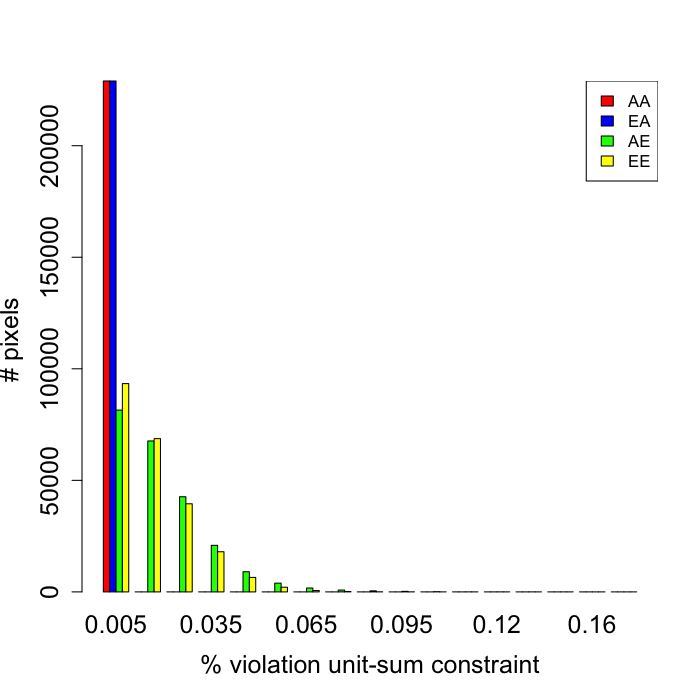}c)
    \caption{Synthetic data results: a) boxplots of the sample mean error. Blue points and red segments are respectively the spatial mean and spatial standard deviation of the sample mean error. b) the mean (across realizations) number of pixels that violate the positivity and  unit-sum constraint for the four methods considered. c) relative violation of the unit-sum constraint. A value of $1\%$ on the x-axis indicates that the reconstructed psfs sums e.g. to 1.01.}
    \label{fig:normbehaviour}
\end{figure}

For each method, we consider a set of $100$ realizations of the fine scale compositional field, each yielding a reconstructed field after the upscaling-down\-scaling process. The upscaling factor $P$ is set each time by randomly and independently sampling in the discrete range $\{2^2,3^2,...,30^2\}$. For each method and each realization we compute the sample mean error, i.e. the average of the Euclidean distance between initial and reconstructed psfs fields -- the average being taken over the realizations. The sample mean error is not computed through the Aitchison distance as this is not defined for recontructed psfs violating the compositional constraints.

Even if the distribution of the sample mean error between the considered methods, reported in Figure~\ref{fig:normbehaviour} a), would suggest a substantial equivalence among the methods, Figures~\ref{fig:normbehaviour} b), c) clearly show that, unlike ATPRCoK, ILR-ATPRCoK is able to produce psfs maps that are consistent with the unit-sum and positivity constraints. Indeed, the ATPRCoK shows a violation of the positivity constraint in about $10^3$ pixels on average, representing roughly $0.5\%$ of the study area.
\medskip

\begin{figure}[htbp]
    \centering
    \includegraphics[scale=.25]{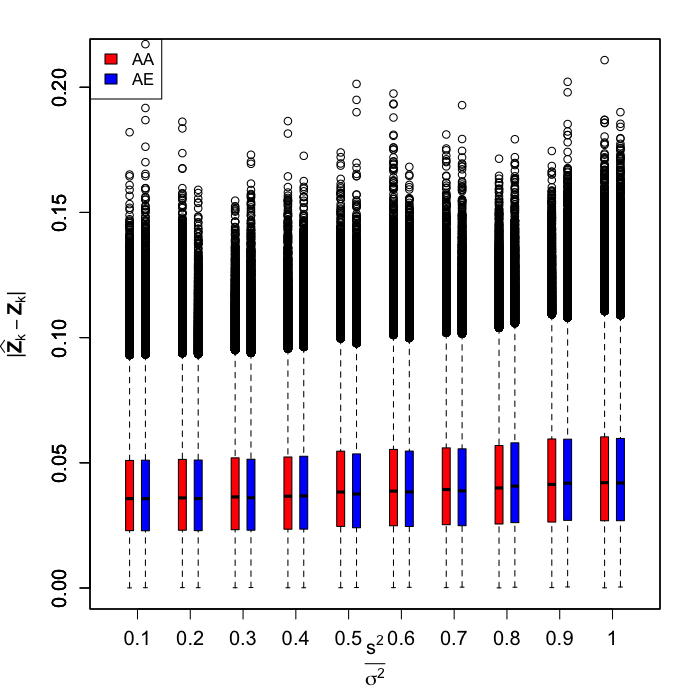}a)\hfill
    \includegraphics[scale=.25]{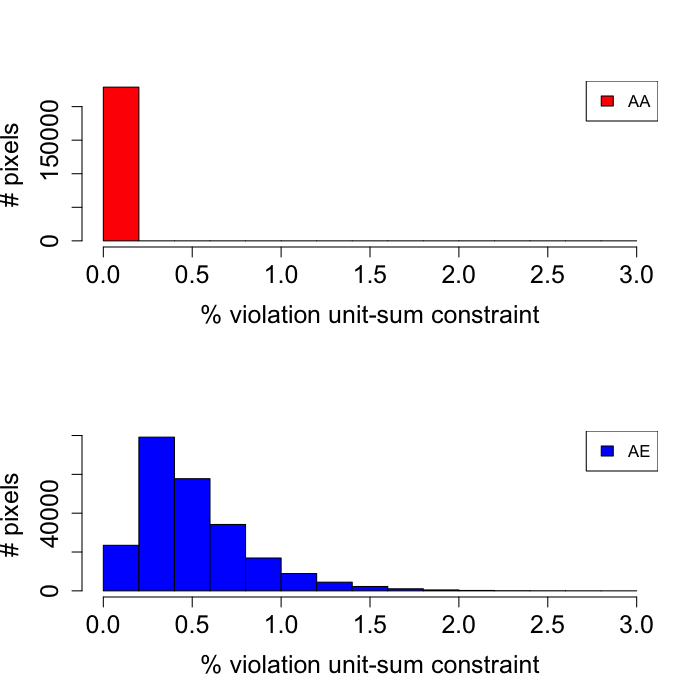}c)\vfill
    \includegraphics[scale=.25]{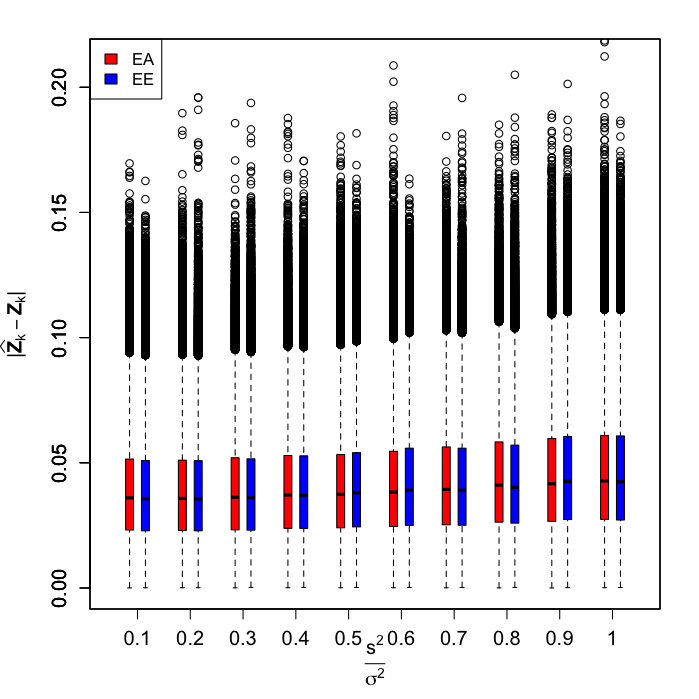}b)\hfill
    \includegraphics[scale=.25]{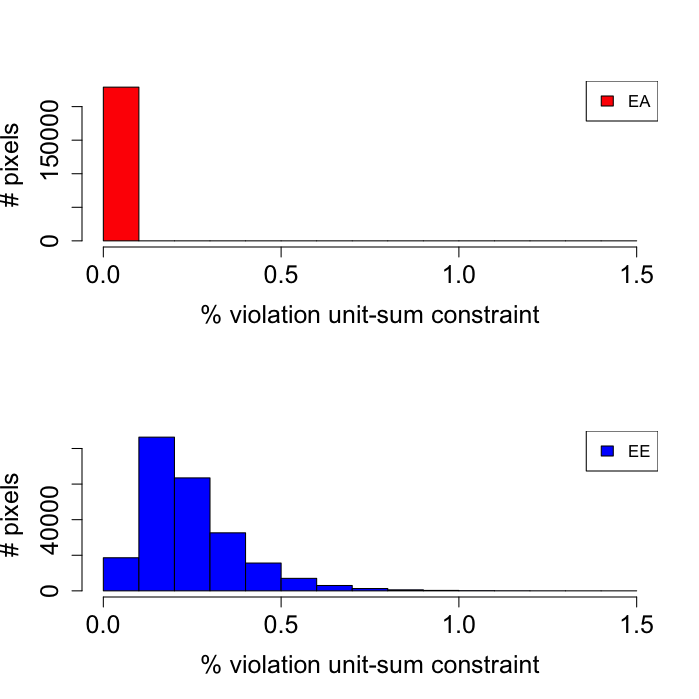}d)
    \caption{Validation on synthetic data. a), b) Boxplots of the error maps between initial and predicted psfs data. c), d) Histograms of the maximum of the violation maps of the unit-sum constraint experienced during the tested variances $s^2$. In panels c), d) the histograms corresponding to AA and EA are fully concentrated on the value 0.}
    \label{fig:realSoilGridspiovernaIdealized}
\end{figure}

We then perform a sensitivity analysis with respect to random perturbations of the initial data of the downscaling procedure, for the four methods described above. This case is representative of input data characterized by a given degree of uncertainty. We thus consider a realization of the synthetic psfs field in case of $\boldsymbol{\mu} = (\frac{1}{3},\frac{1}{3},\frac{1}{3})$ and sill $\sigma^2=0.1$ (i.e. as in Figure~\ref{fig:idealizedTriangle}) and we set the upscaling factor to $P=225$. Let us indicate with $K=1,\dots,M$ the elements of the coarse maps. The upscaled maps $\mathbf{Z}_K$ are then perturbed with a set of i.i.d. Gaussian random errors $\boldsymbol{\epsilon}_K$. Similarly as before, these perturbations were generated on the ILR transforms, by adding a zero-mean independent Gaussian error with variance $s^2,$  ranging from $10\%$ to $100\%$ of the sill $\sigma^2$. 

In Figure~\ref{fig:realSoilGridspiovernaIdealized} a), b) we report the boxplots of the error maps for each value of $s^2$. The error maps are computed, for each pixel, as the Euclidean distance between initial and predicted psfs. We note that both ATPRCoK and ILR-ATPRCoK are quite robust even in case of relatively high perturbations of the initial data. In Figure~\ref{fig:realSoilGridspiovernaIdealized} c), d) we show the histograms of the maximum, across simulations, of the violation maps of the unit-sum constraint. For instance, the vertical bar in correspondence of the range [1,1.2] in Figure~\ref{fig:realSoilGridspiovernaIdealized} d) indicates the count of pixels whose maximum discrepancy (across simulations) from unity of the sum of psfs is between 1\% and 1.2\% (i.e., the sum is in [1.01,1.012] or [0.988,0.99]). These results clearly show the ability of ILR-ATPRCoK method to produce results consistent with the unit-sum constraint, unlike the ATPRCoK method which yields maps with a significant violation of the aforementioned constraint.

\subsection{Downscaling SoilGrids data}
In this section, we test the performances of the proposed method in downscaling psfs from soil digital maps publicly available. This case is considered to analyse compositional random fields having a realistic spatial distribution. For this purpose and in view of our case study, we consider SoilGrids, which is a system for automated digital soil mapping based on state-of-the-art spatial predictions methods~\cite{hengl:2014, hengl:2017} released in 2014 by ISRIC (International Soil Reference and Information Centre) - World Soil Information, a non-profit organization funded by the Dutch government. 
SoilGrids predictions are based on globally fitted models using soil profile and environmental covariate data.
When first released, SoilGrids provided a collection of soil properties and class maps of the world at $1 \, km$ spatial resolutions, produced using automated soil mapping based on statistical regression models.
In $2017$, the resolution has been increased to $ 250 \, m$ and the accuracy of the predictions has been greatly improved by using machine learning algorithms instead of the previously employed linear regression~\cite{hengl:2017}. In $2020$, SoilGrids released a version where, among other updates, soil map predictions are provided with a mean value together with an uncertainty level map. 
SoilGrids data are available publicly under the Open DataBase License. Among SoilGrids predicted variables, relevant to this work are clay, silt and sand percentages at different soil depths. In this section, the values considered for geostatistical downscaling are those referred to the topsoil, i.e. depth of $0 \, cm$. 

\begin{figure}[htbp]
    \centering
    \includegraphics[scale=.25]{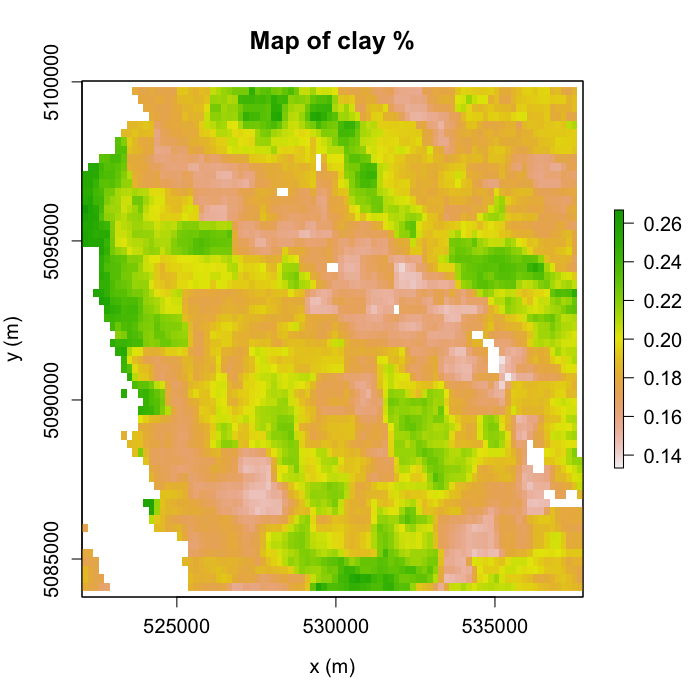}\hfill
    \includegraphics[scale=.25]{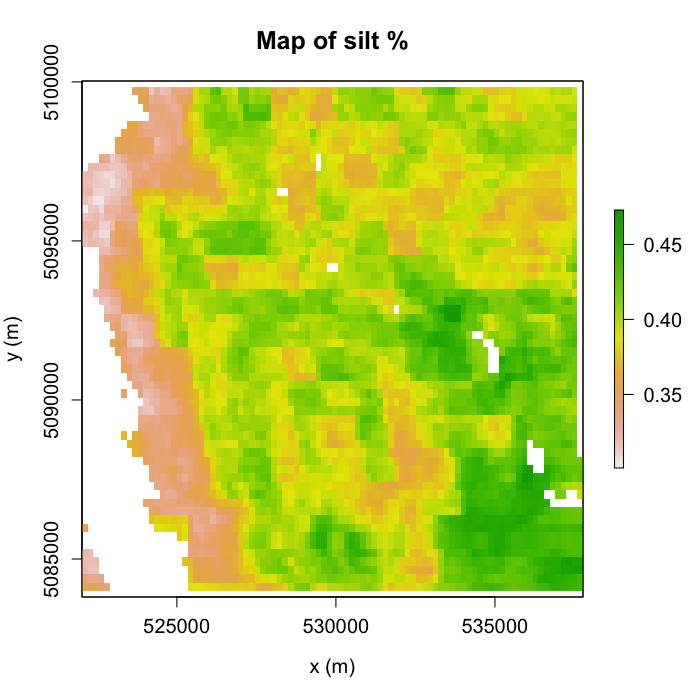}\vfill
    \includegraphics[scale=.25]{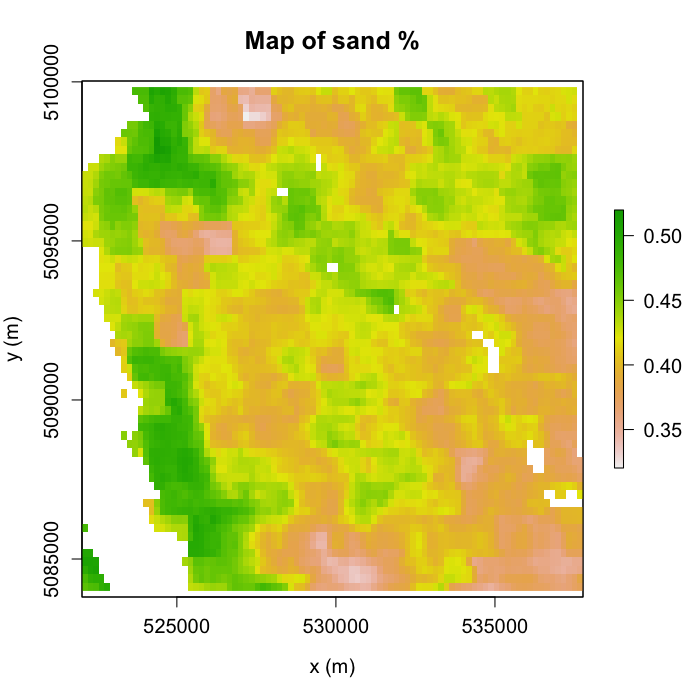}\hfill
    \includegraphics[scale=.25]{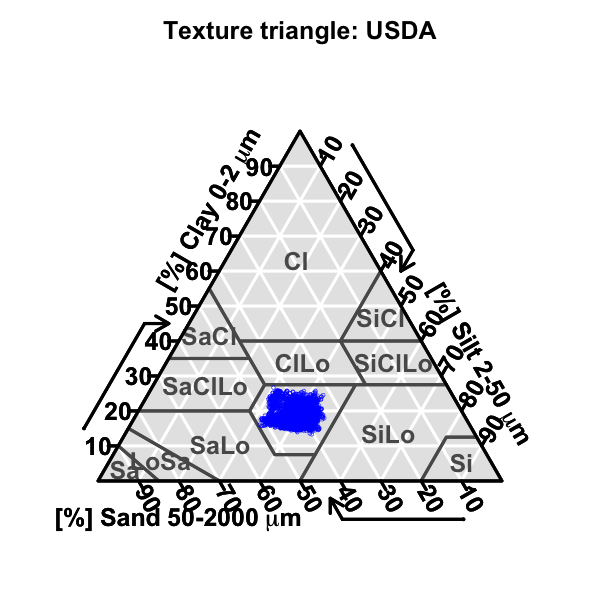}
    \caption{SoilGrids psfs data within the study region. The considered data refer to the mean value of psfs as reconstructed in the SoilGrids repository.}
    \label{fig:idealizedTrianglerealpioverna}
\end{figure}

\begin{figure}[htbp]
    \centering
    \includegraphics[scale=.25]{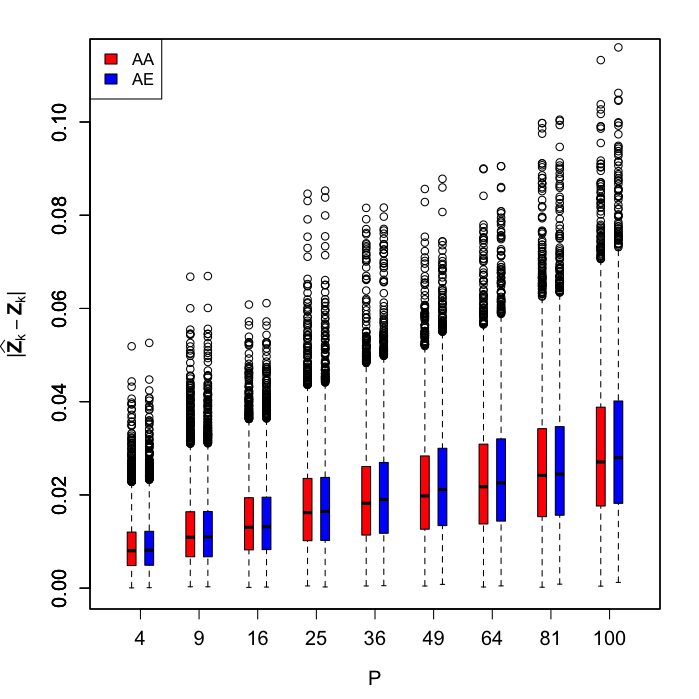}a)\hfill
    \includegraphics[scale=.25]{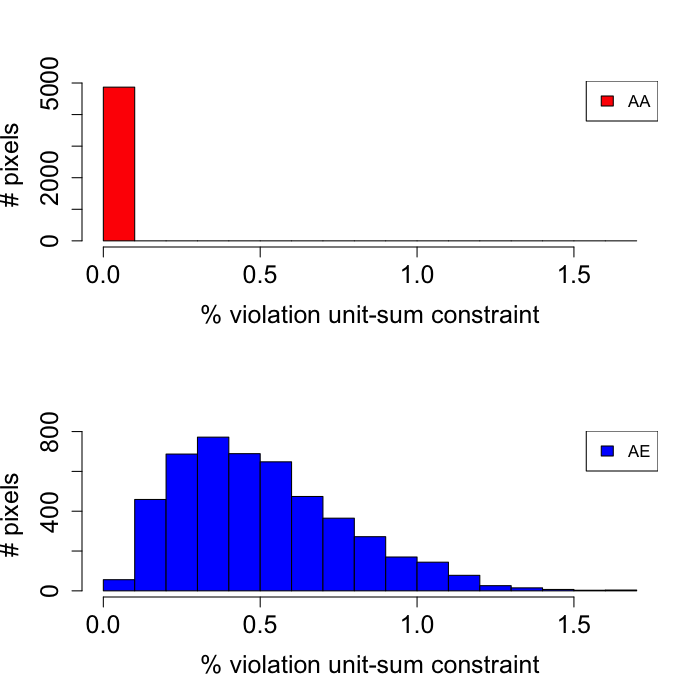}c)\vfill
    \includegraphics[scale=.25]{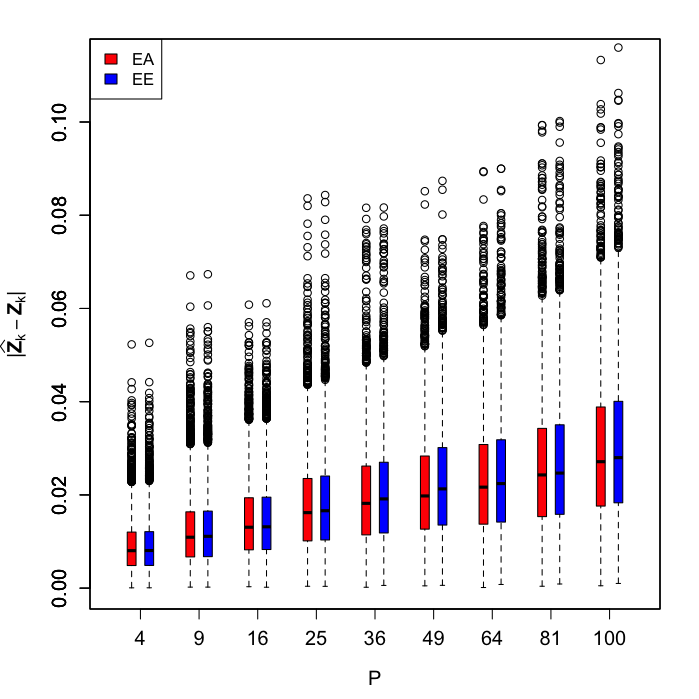}b)\hfill
    \includegraphics[scale=.25]{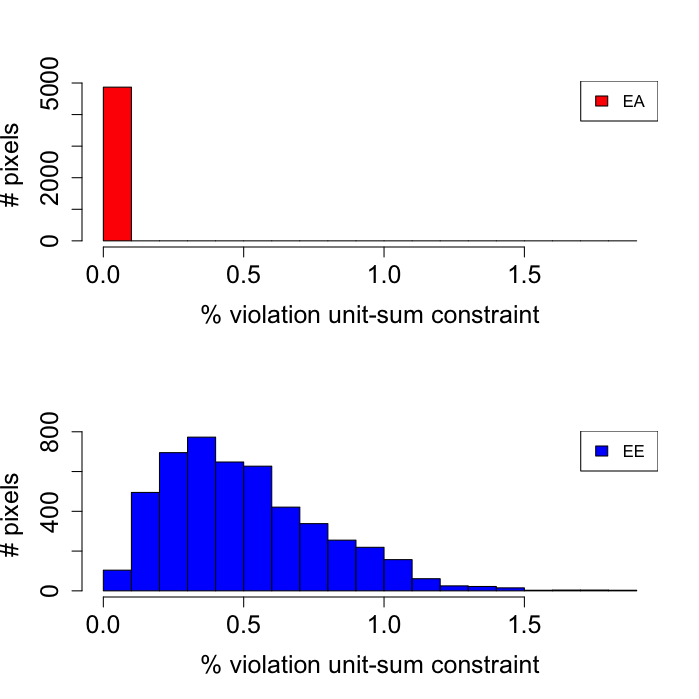}d)
    \caption{Downscaling on SoilGrids data. a), b) Boxplots of the error maps between SoilGrids and predicted psfs data. c), d) Histograms of the maximum of the violation maps of the unit-sum constraint experienced during the tested upscaling factors. In panels c), d) the histograms corresponding to AA and EA are fully concentrated on the value 0.
    }
    \label{fig:realSoilGridspioverna}
\end{figure}

We focus on a geographical domain with area $|D| = 15750\times16000\:m^2,$ located in a pre-Alpine area, more specifically the basin of Pioverna river in the Lombardy region in Northern Italy near the city of Lecco. This region was selected as it is  similar, from the geomorphological viewpoint, to the area analyzed in the case study presented in Section~\ref{sec:CaldoneCaseStudy}. The psfs as available in SoilGrids are reported in Figure~\ref{fig:idealizedTrianglerealpioverna}.
Based on these data, we test the performance of the ILR-ATPRCoK method, at different levels of the upscaling factor $P$. 
Following the procedure described in Subsection~\ref{SyntheticDataAnalysis}, we consider a sequence of upscaling-down\-scaling operations, both in Aitchison and Euclidean geometry, of the SoilGrids data in Figure ~\ref{fig:idealizedTrianglerealpioverna}. For each upscaling factor $P$ in the range $\{2^2,3^2,...,10^2\}$ and for each pixel in $D$, we compute the Euclidean distance of the psfs estimates from the initial SoilGrids data, yielding a set of error maps (one for each value of $P$). These are displayed through boxplots in Figure~\ref{fig:realSoilGridspioverna} a), b). We note that, mainly at high uspcaling factors, the ILR-ATPRCoK method shows slightly better behaviour with respect to the classical ATPRCoK, producing solutions that are closer to the reference ones w.r.t the results produced via ATPRCoK downscaling. 

In Figures~\ref{fig:realSoilGridspioverna} c), d), we report the histograms of the maximum of the violation maps of the unit-sum constraint experienced during the set of upscaling factors, for the four cases being considered. Interpretation of these histograms is fully analogous to that in Figure \ref{fig:realSoilGridspiovernaIdealized}. These results confirm that the ILR-ATPRCoK method is able to produce downscaled maps that are consistent with the unit-sum constraint, as opposed to the ATPRCoK downscaling method. Finally, we do not report any violation of the positivity constraint for ATPRCoK, differently from what shown in the tests reported in Section~\ref{SyntheticDataAnalysis}.

\section{A case study}\label{sec:CaldoneCaseStudy}
Our case study considers an application to a domain $D$ centered on the city of Lecco, located in the Lombardy region in  Northern Italy, which is crossed by three streams (Bione, Caldone, and Gerenzone) that have the typical characteristics of the pre-Alpine area. The hydrographic basin of the Caldone water course is $24\:km^2$ wide, with an altitude ranging from $197\,m$ a.m.s.l. to $2118\,m$ a.m.s.l. at the top of Grigna Meridionale mountain. Geologically, the basin is characterized by rocky outcrops in the higher part (mainly limestone and clastic rock), while downstream towards the city the river flows through a floodplain. The average precipitation over the city of Lecco is about $1400\:mm/yr$. 
\begin{figure}[htbp]
  \centering
  \includegraphics[scale = 0.3]{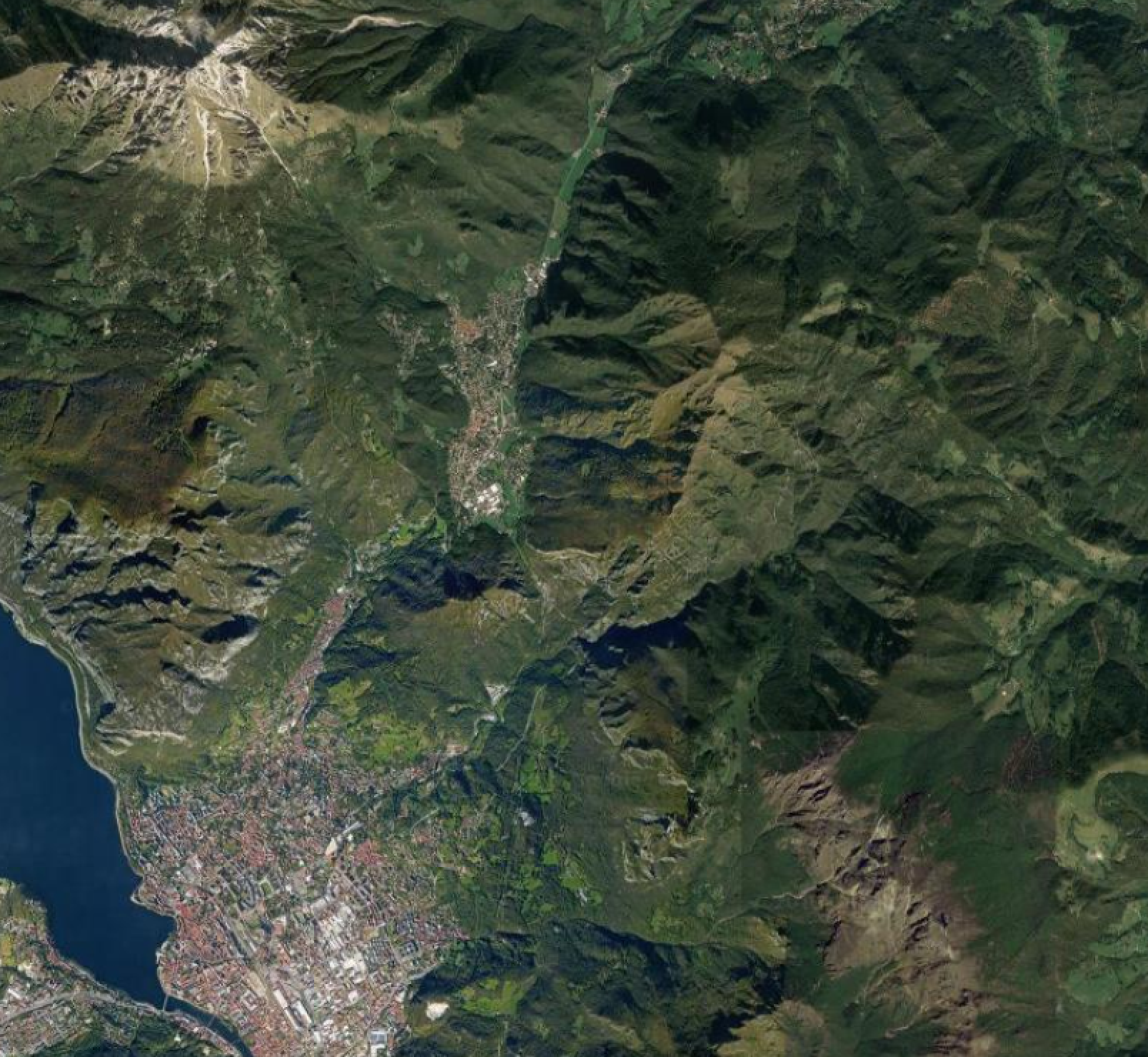}
  \caption{Aerial view of the case study area.}
  \label{fig:aerial}
\end{figure}

The combination between a short hydrologic response time, high slope, intense sediment transport and flow within a densely urban area makes the Caldone river a suitable case study for hydrogeological instability and hazard. 
This motivates the development of numerical models intended to simulate hydrogeological processes, such as the SMART-SED simulation tool (Sustainable Management of sediment transpoRT in responSE to climate change conDitions)~\cite{WLF5Book} which is able to simulate sediment transport resulting from slope erosion. These models typically need to be initialized with psfs maps, with a resolution consistent with the Digital Terrain Model (DTM), to be able to model properly the hydrological processes taking place in the study region.
However, field measurements of psfs are not available at the study site, which motivates the use of public repository to obtain indirect information on these input data.

\begin{figure}[htbp]
    \centering
    \includegraphics[scale=.25]{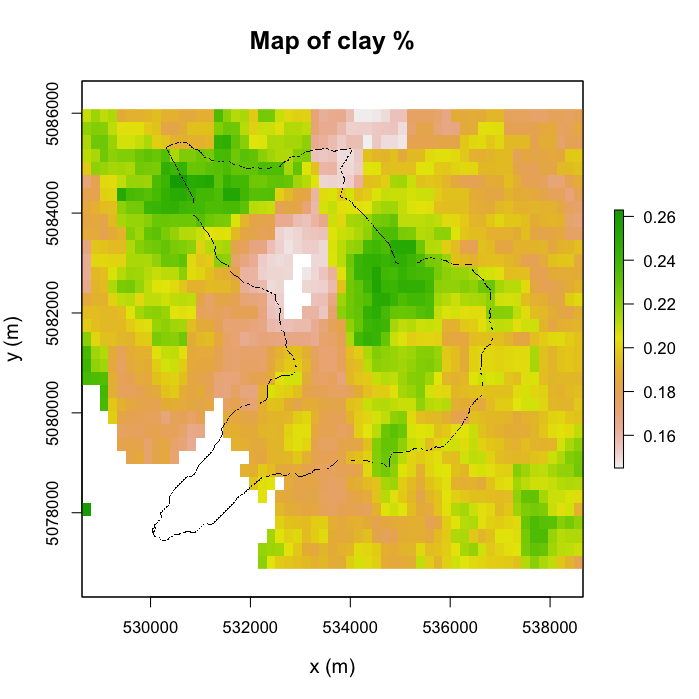} \hfill
    \includegraphics[scale=.25]{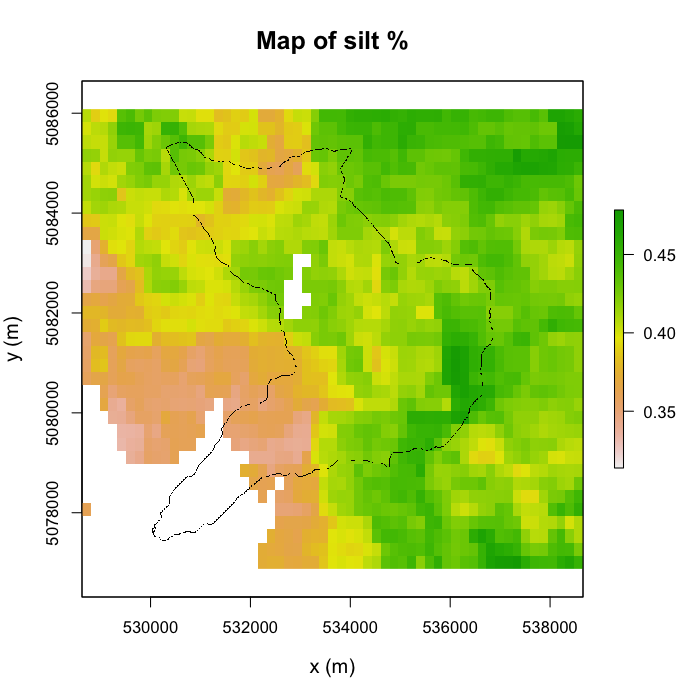} \vfill
    \includegraphics[scale=.25]{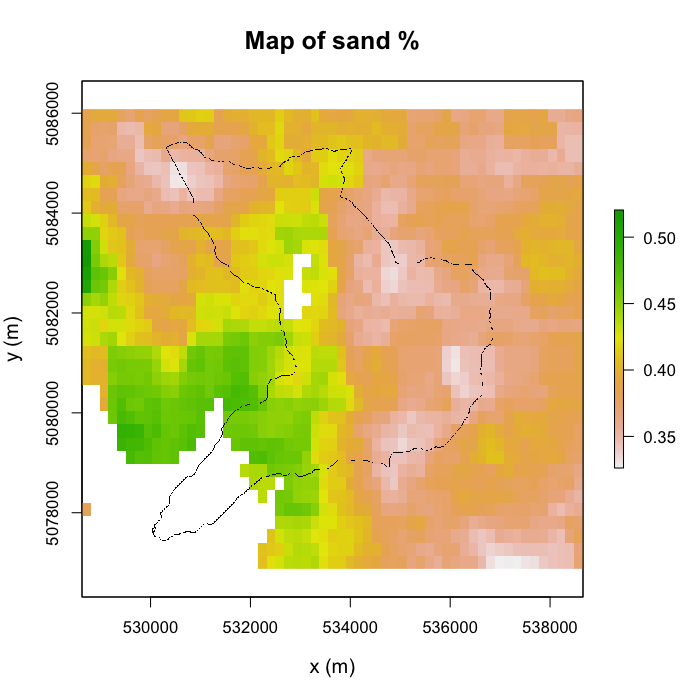} \hfill
    \includegraphics[scale=.25]{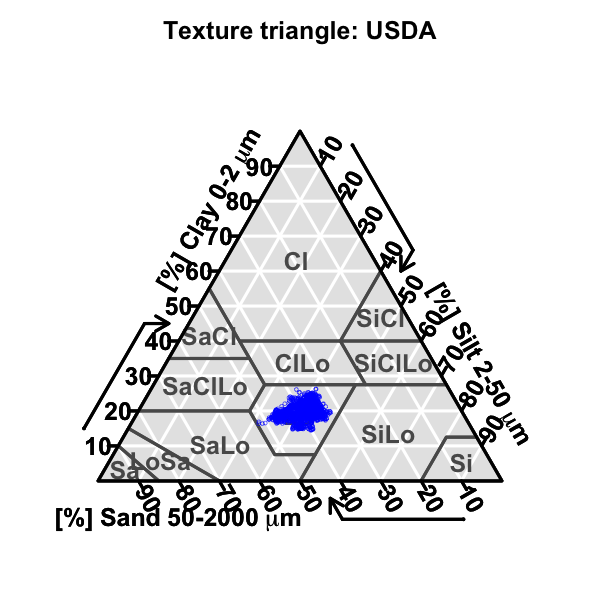}
    \caption{Clay, silt and sand maps coming from SoilGrids. In black is shown the polygon delimiting the hydrographic Caldone basin.}
    \label{fig:psf_maps}
\end{figure}

SoilGrids psfs at the study region have a pixel support with measure $|\nu_K| = 250\:\times\:250 \:m^2$. In terms of the USDA classification, the soil texture of the SoilGrids data for the present case study falls into the loam category. This kind of soil texture, according to~\cite{rosso:2004}, is classified as fairly permeable soil with moderate infiltration rates and moderate runoff potential. In the following, these coarse-scale data are downscaled to the resolution of the DTM employed for the SMART-SED model, i.e. $5\:m$, using ILR-ATPRCoK, following the methodology described in Section~\ref{sec:CompositionalATPRK}. Together with ILR-ATPRCoK results, we here aim to provide random realizations of the psfs fields -- obtained via Block Sequential Gaussian Simulation (BSGS) -- as demonstration of the ability of the method to produce stochastic compositional maps compatible with coarse scale data. 

\begin{figure}
  \centering
  \includegraphics[scale=.5]{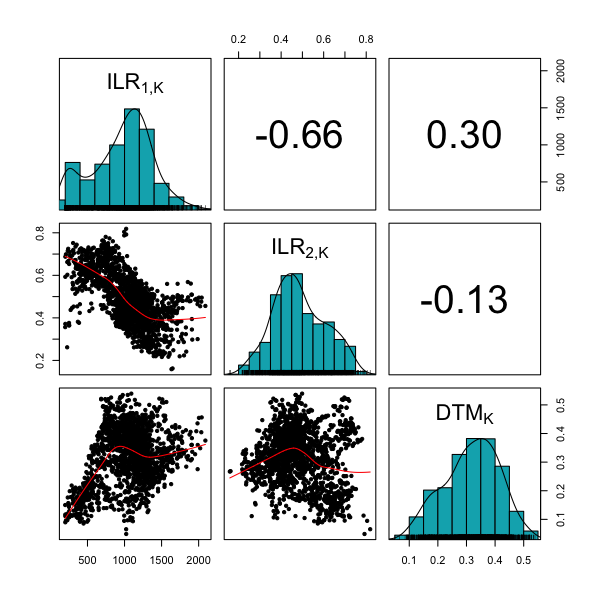}
  \caption{Scatter plots, histograms and Pearson coefficients of the ILRs and the DTM at coarse resolution.}
  \label{fig:pearsonDTMILR}
  \vspace{0.5cm}
\end{figure}

\begin{figure}
  \centering
  \includegraphics[width=\textwidth]{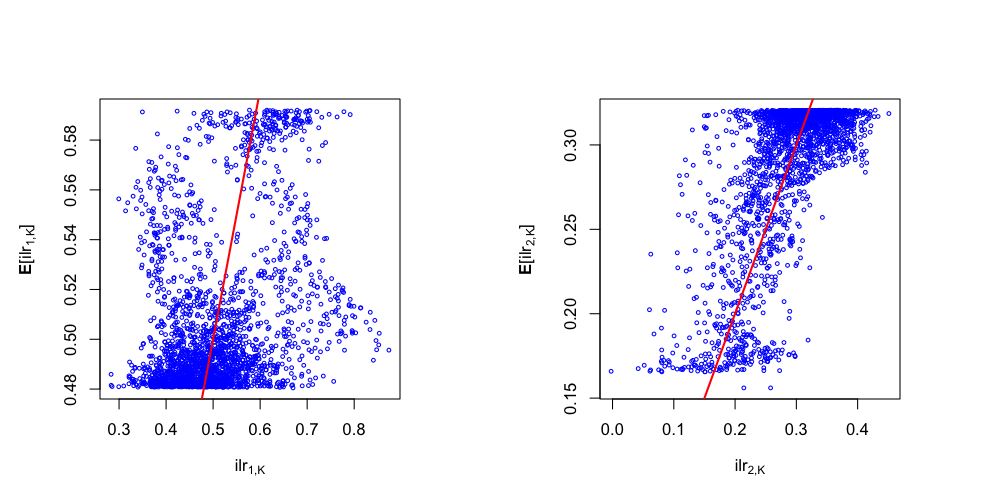}
  \caption{Scatter plots of the observed values and the fitted values of the regression model. In red, the line of equation $\mathbb{E}[\widehat{\text{ILR}}_{i,K}] = \text{ILR}_{i,K}, i =1,2$. The Pearson coefficient is $0.67$ for $\mathbb{E}[\widehat{\text{ILR}}_{1,K}], \text{ILR}_{1,K}$ and $0.43$ for $\mathbb{E}[\widehat{\text{ILR}}_{2,K}], \text{ILR}_{2,K}$.}
  \label{fig:ILR_regression}
  \vspace{0.5cm}
\end{figure}

Based on SoilGrids psfs data, we define the following coarse resolution maps
\begin{equation}
    (\text{ILR}_{1,K}, \text{ILR}_{2,K})^{'} = \text{ILR}( \ (Z_{1,K}, Z_{2,K}, Z_{3,K})' \ ), \ \ K=1,\dots,M.
\end{equation}
In the ILR-ATPRCoK model, we consider as covariates $u_k^l, l=1,\dots,L$, the DTM and its square, driven by the parabolic relation displayed in the scatterplot in Figure~\ref{fig:pearsonDTMILR}. 
For the fine map predictions $\widehat{\text{ILR}}_{1,k}$, $\widehat{\text{ILR}}_{2,k}$ we thus consider the model
\begin{align}
    \begin{split}
        \widehat{\text{ILR}}_{1,k} &=  \beta_0^{(1)} + \beta_1^{(1)} \cdot \text{DTM$_k$} + \beta_2^{(1)} \cdot \text{DTM$_k^2$} + \sum_K \lambda_K e_{1,K}, \\
        \widehat{\text{ILR}}_{2,k} &=  \beta_0^{(2)} + \beta_1^{(2)} \cdot \text{DTM$_k$} + \beta_2^{(2)} \cdot \text{DTM$_k^2$} + \sum_K \lambda_K e_{2,K}.
    \end{split}
    \label{models}
\end{align}
The fitted values are plotted against the observed values in Figure~\ref{fig:ILR_regression}. 

\begin{figure}
  \centering
  \includegraphics[scale=.30]{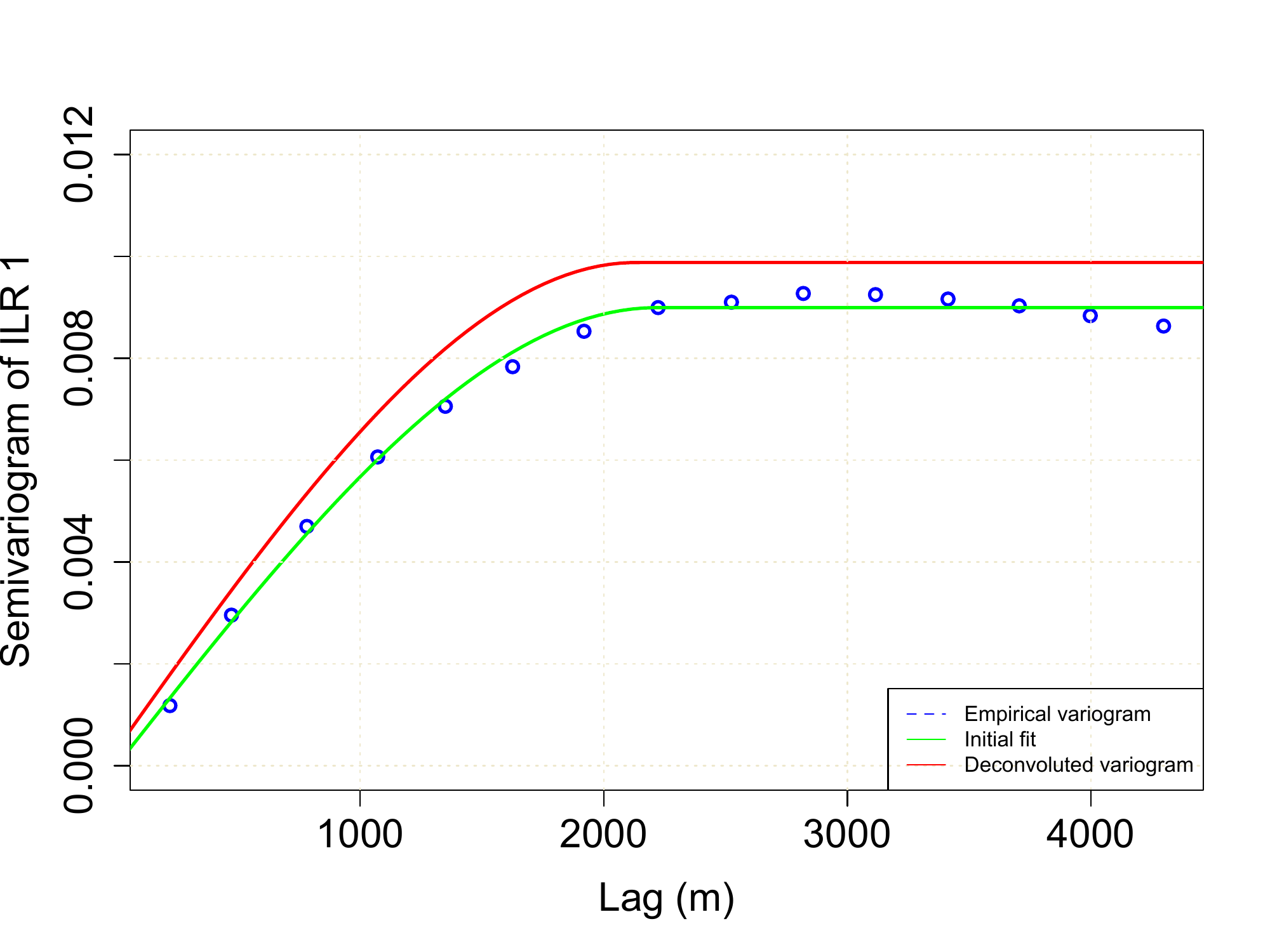}a) 
  \includegraphics[scale=.30]{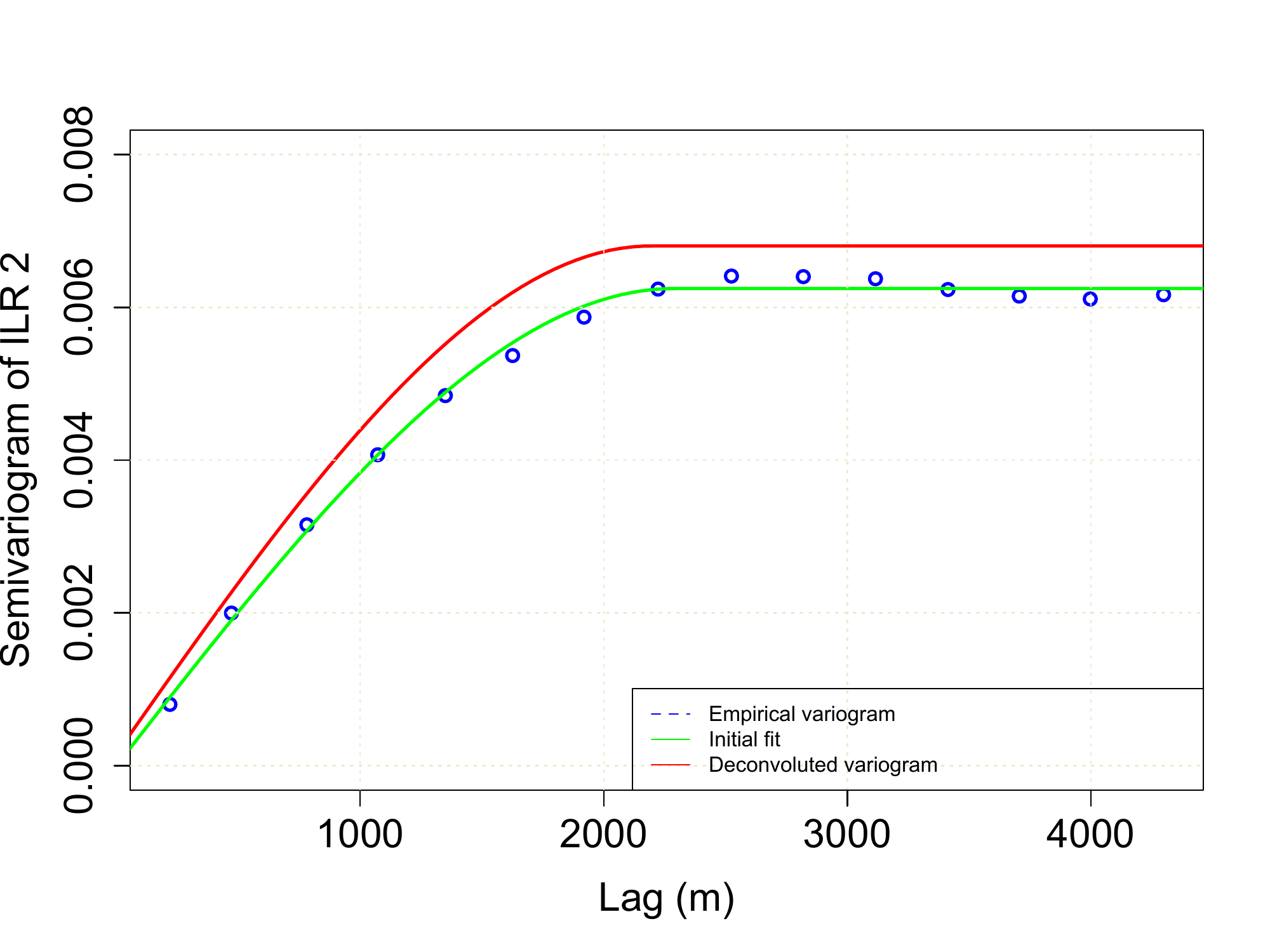}b)
  \caption{Results of the Goovaerts’ deconvolution (red line) procedure applied to the empirical variograms (blue dots) starting from an initial fit (green line). The empirical variograms are fitted to a spherical variogram model. Fitted models: (a) Sill: 0.00956, Range: 2130 $m$, Nugget: 0.00032; (b) Sill: 0.00665, Range: 2190 $m$, Nugget: 0.00016.}
  \label{fig:Goovaerts}
  \vspace{0.5cm}
\end{figure}
\begin{figure}
    \centering
    \includegraphics[width=\textwidth]{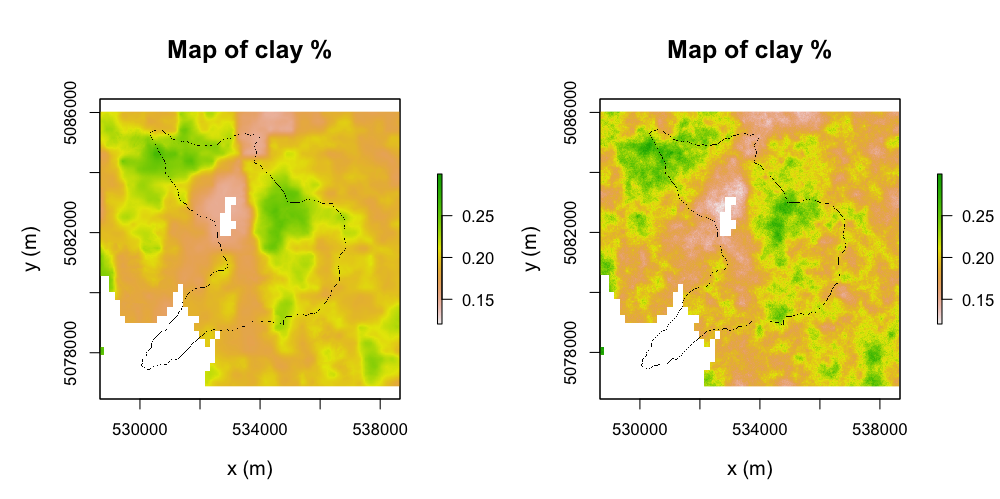} \\
    \includegraphics[width=\textwidth]{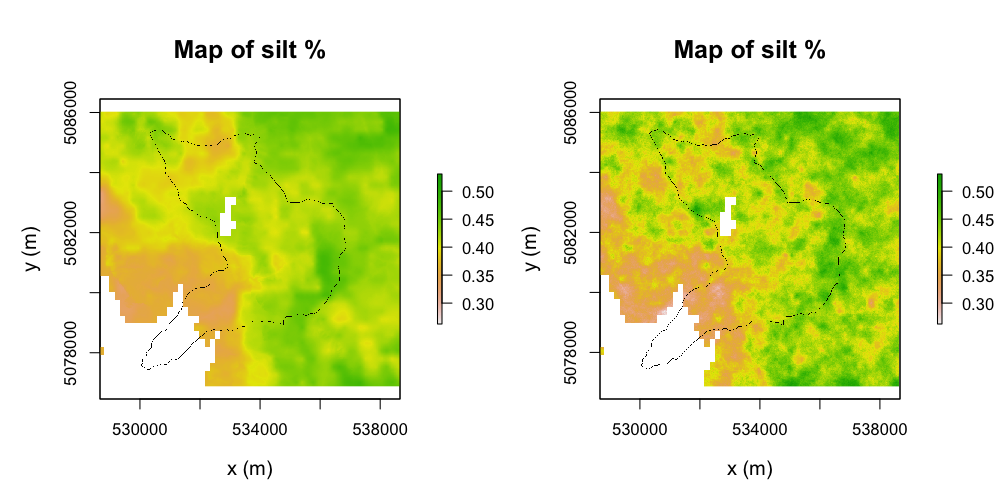} \\
    \includegraphics[width=\textwidth]{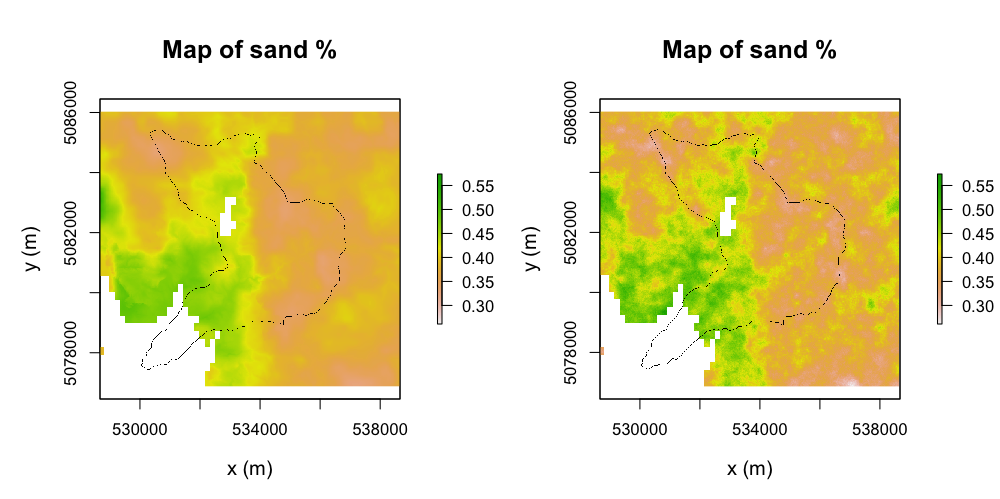} 
    \caption{Clay, silt and sand maps results of the ILR-ATPRCoK on the left and on the right results of BSGS. The black polygon delimits the hydrographic Caldone basin.}
    \label{fig:psf_maps_downscaled}
\end{figure}

To perform ILR-ATPK, the spatial correlation structure of the fine residuals $e_{1,k}$ and $e_{2,k}$ is estimated by applying the Goovaerts’ deconvolution procedure to the variograms fitted to the coarse residuals $e_{1,K}$ and $e_{2,K}$ (based on a spherical model with nugget), and by assuming $e_{1,k}$ and $e_{2,k}$ to be uncorrelated. The latter assumption is supported by the residuals' analysis (not reported for brevity). Once the fine variograms of the residuals have been estimated, it is possible to solve the ATPK linear system, according to \eqref{kriging_system}. The downscaled ILR are then backtransformed in the Aitchison space in order to get downscaled psfs, see Figure~\ref{fig:psf_maps_downscaled}, left column. 

Finally, in Figure~\ref{fig:psf_maps_downscaled}, right column, we show a sample realization for the downscaled psfs, obtained via BSGS. These stochastic maps shall form the cornerstone to evaluate the propagation of the uncertainty associated with the psfs through the SMART-SED model, and eventually assess the sensitivity of the sediment transport model to this information.

\section{Conclusions}\label{sec:conclusion}
We have presented a novel downscaling method for compositional data, based on the ATPRCoK method in the Aitchison geometry, with application to the geostatistical downscaling of psfs data. We have tested the method first in the case of synthetic data and then on a dataset from the SoilGrids online repository. In particular, we have shown the ability of the method to automatically handle  the compositional nature of the considered data. Indeed, the proposed method produces maps that respect the unit-sum and positivity constraints, as opposed to the classical ATPRCoK method that produces maps which are not consistent with the compositional constraints. 

Validation on both synthetic and SoilGrids data show good performances of the method in downscaling, as well as robustness to the uncertainty of the input data. This is critical to the use of data from public repositories in local analyses, when point observations are not available, as they are naturally prone to uncertainty at a fine scale. While a full account of SoilGrids uncertainty will be the scope of future work, a relevant feature of ILR-ATPRCoK method -- similarly as ATPRCoK in the Euclidean setting -- is the possibility to easily incorporate point observations collected at the site, thus anchoring the downscaled maps (either kriged or simulated) to such observations~\cite{park:2013}. For instance, at the time of writing, a campaign of data acquisition is under way in the Caldone basin, and will support the definition of (possibly improved) random psfs maps, to be used as input to the SMART-SED model discussed in Section \ref{sec:CaldoneCaseStudy}. 

\section*{Acknowledgements}
The authors gratefully acknowledge the financial support of Fondazione Cariplo in the framework of the SMART-SED project, grant number 2017-0722.

\begin{appendices}

\section{Proof of ILR-ATPRCoK proposition}\label{appendixProof}

In the following we propose a proof of Prop.~\ref{PropositionNewComp}, i.e. the equivalence of the ILR-ATPRCoK predictor (in the simplex $\mathbb{S}^p$) to the classical ATPRCoK predictor (in the space $\mathbb{R}^{p-1}$) by applying an isometric isomorphism. The equivalence must be intended in the predictor, unbiasedness and optimality conditions.
In the following we make extensive use of the ILR properties defined in~\cite{pawlowsky2015modeling} (p. 37-43) and the fact that $ILR \colon \mathbb{S}^p \to \mathbb{R}^{p-1}$ extract the Fourier coordinates of a basis projection for the vector $\mathbf{z} \in \mathbb{S}^p$, i.e.,
\begin{align*}
 \mathbf{y} &= ILR(\mathbf{z}) \\
    \mathbf{z} &= \bigoplus_{i=1}^{p-1} \langle\mathbf{z}, \boldsymbol{\psi}_i\rangle_a \odot \boldsymbol{\psi}_i = (\mathbf{y}'\odot\Psi)^{'} = \mathcal{C}\left[\prod_{i=1}^{p-1}\psi_{i,1}^{y_i},\dots,\prod_{i=1}^{p-1}\psi_{i,p}^{y_i}\right] = ILR^{-1}(\mathbf{y}), 
\end{align*}
where the rows of $\Psi = [\psi_{i,j}]_{i=1,\dots,p-1}^{j=1\dots,p}$ are (compositional) vectors identifying an orthonormal basis of the simplex $\{\boldsymbol{\psi}_1,\dots,\boldsymbol{\psi}_{p-1}\}$ and $\mathbf{y}=[y_i]_{i=1,\dots,p-1} \in \mathbb{R}^{p-1}$ is the vector of coordinates (i.e. of the Fourier coefficients) of the identified basis of the simplex. 

Let us start with the predictor, applying the ILR to the ATPRCoK predictor defined in the Aitchison space $\mathbb{S}^{p}$~\eqref{area_to_point_kriging_estimate_aitchison}, we get
\begin{equation}
    \widehat{\mathbf{Y}}_k = \sum_l u_k^l \boldsymbol{\beta}^l_{\mathbf{Y}} + \sum_K ILR(\Lambda_K \boxdot ((\mathbf{e}_K^{\mathbf{Y}})' \odot \Psi)^{'}).
\end{equation}

\noindent where $\boldsymbol{\beta}^l = ILR^{-1}(\boldsymbol{\beta}^l_{\mathbf{Y}})$ and $\mathbf{e}_K = ((\mathbf{e}_K^{\mathbf{Y}})'\odot \Psi)^{'}.$ Being $e_{K,i}^{\mathbf{Y}},$ the $i$-th element of the vector $\mathbf{e}_K^{\mathbf{Y}}$, $i=1,...,p-1$, we have 
\begin{align*}
ILR(\Lambda_K \boxdot ((\mathbf{e}_K^{\mathbf{Y}})'\odot \Psi)^{'}) &= ILR(((\mathbf{e}_K^{\mathbf{Y}})'\odot \Psi \boxdot \Lambda_K^{'})^{'}) = \\
&= ILR(((\mathbf{e}_K^{\mathbf{Y}})' \odot [\boldsymbol{\psi}_i^{'} \boxdot \Lambda_K^{'}]_{i=1,\dots,p-1})^{'}) = \\
&= ILR\left(\left(\bigoplus_{i=1}^{p-1} e^{\mathbf{Y}}_{K,i} \odot \boldsymbol{\psi}_i^{'} \boxdot \Lambda_K^{'}\right)^{'}\right) = \\
&= \left(\sum_{i=1}^{p-1} e^{\mathbf{Y}}_{K,i} ILR(\boldsymbol{\psi}_i^{'} \boxdot \Lambda_K^{'})\right)^{'} = \\
&= ((\mathbf{e}_K^{\mathbf{Y}})^{'} [ILR(\boldsymbol{\psi}_i^{'} \boxdot \Lambda_K^{'})]_{i=1,\dots,p-1})^{'} = \\
&= [\langle\Lambda_K \boxdot \boldsymbol{\psi}_i, \boldsymbol{\psi}_j\rangle_a]_{i,j=1,\dots,p-1} \mathbf{e}_K^{\mathbf{Y}} = \Lambda_K^{\mathbf{Y}} \mathbf{e}_K^{\mathbf{Y}}.
\end{align*}

In this way, we obtain the ATPRCoK predictor in the Euclidean space $\mathbb{R}^{p-1},$ i.e.,
\begin{equation}
    \widehat{\mathbf{Y}}_k = \sum_l u_k^l \boldsymbol{\beta}^l_{\mathbf{Y}} + \sum_K \Lambda_K^{\mathbf{Y}} \mathbf{e}_K^{\mathbf{Y}}.
\end{equation}
Regarding the unbiasedness and optimality conditions, calling $\overline{\boldsymbol{\mu}}^{\mathbf{Y}} = ILR(\overline{\boldsymbol{\mu}}),$ $d(\cdot,\cdot)$ the Euclidean distance and considering ILR properties, one easily obtains that the same conditions hold in $\mathbb{R}^{p-1},$
\begin{align*}
&\mathbb{E}\left[d^2_a\left(\bigoplus_K\Lambda_K\boxdot\mathbf{e}_K, \mathbf{e}_k\right)\right] = \mathbb{E}\left[d^2\left(
\sum_K\Lambda_K^\mathbf{Y}\mathbf{e}^\mathbf{Y}_K, \mathbf{e}^\mathbf{Y}_k\right)\right], \\
&\text{ILR} \left( Cen\left(\bigoplus_K\Lambda_K\boxdot\mathbf{e}_K\right) \right) = \mathbb{E}\left[ILR\left(\bigoplus_K\Lambda_K\boxdot\mathbf{e}_K\right)\right]  = \mathbb{E}\left[\sum_K\Lambda_K^\mathbf{Y}\mathbf{e}^\mathbf{Y}_K\right] = \overline{\boldsymbol{\mu}}^{\mathbf{Y}}.
\end{align*}

Finally from an ``energy'' point of view the two formulation are equivalent. Indeed if we consider the quadratic form associated with the covariance structure $C_a(\mathbf{x}_1,\mathbf{x}_2)$, $\mathbf{x}_1,\mathbf{x}_2 \in D$, denote by $\mathbf{z}$ a non-random element of $\mathbb{S}^p$, and use the ILR properties, we obtain
\begin{align*}
\xi &= \langle \mathbf{Z}(\mathbf{x}_1) \ominus \boldsymbol{\mu}(\mathbf{x}_1), \mathbf{z} \rangle_a = \langle ILR(\mathbf{Z}(\mathbf{x}_1) \ominus \boldsymbol{\mu}(\mathbf{x}_1)), \mathbf{y} \rangle =\\
& = \langle\mathbf{Y}(\mathbf{x}_1) - \boldsymbol{\mu}_{\mathbf{Y}}(\mathbf{x}_1), \mathbf{y} \rangle.
\end{align*}
Using the latter expression, one has
\begin{align*}
    \langle C_{a}(\mathbf{x}_1,\mathbf{x}_2)\mathbf{z}, \mathbf{z}\rangle_a &= \langle ILR(C_a(\mathbf{x}_1,\mathbf{x}_2) \mathbf{z}), \mathbf{y}\rangle = \\
 & =  \langle \mathbb{E}[\xi\, ILR(\mathbf{Z}(\mathbf{x}_2) \ominus \boldsymbol{\mu}(\mathbf{x}_2)], \mathbf{y} \rangle = \langle \mathbb{E}[\xi\, (\mathbf{Y}(\mathbf{x}_2) - \boldsymbol{\mu}_{\mathbf{Y}}(\mathbf{x}_2))], \mathbf{y} \rangle = \\
 & = \langle C_{\mathbf{Y}}(\mathbf{x}_1,\mathbf{x}_2) \mathbf{y}, \mathbf{y} \rangle.
\end{align*}
Hence, the covariance structure in the Euclidean space $\mathbb{R}^{p-1},$ reads,
$C_{\mathbf{Y}}(\mathbf{x}_1,\mathbf{x}_2) = \mathbb{E}[\langle\mathbf{Y}(\mathbf{x}_1) - \boldsymbol{\mu}_{\mathbf{Y}}(\mathbf{x}_1), \mathbf{y} \rangle \, (\mathbf{Y}(\mathbf{x}_2) - \boldsymbol{\mu}_{\mathbf{Y}}(\mathbf{x}_2))].$
This means that the knowledge of the covariance structure in the Aitchison simplex $C_a$ implies the knowledge of the covariance structure in the Euclidean space $C_{\mathbf{Y}}$ and viceversa. This result, together with the relation stated above among the regression coefficients ($\boldsymbol{\beta}^l = ((\boldsymbol{\beta}^l_{\mathbf{Y}})'\odot\Psi)^{'}$), and the equivalence of predictor, optimality and unbiasedness conditions, are sufficient to prove Proposition~\ref{PropositionNewComp}.

\end{appendices}

\cleardoublepage
\addcontentsline{toc}{chapter}{Bibliography}
\hyphenpenalty=10 %

\bibliographystyle{plain}

\bibliography{MOXReport}

\end{document}